\documentclass[longauth]{aa}

\usepackage{natbib}
\bibpunct{(}{)}{;}{a}{}{,}

\usepackage{graphicx}
\usepackage{txfonts}
\usepackage{csquotes}
\usepackage{booktabs}
\usepackage{amsmath}
\usepackage{amssymb}
\usepackage{amsfonts}
\usepackage{xcolor}
\usepackage{physics}
\usepackage{siunitx}
\AtBeginDocument{\RenewCommandCopy\qty\SI}

\DeclareSIUnit{\parsec}{pc}
\DeclareSIUnit{\pc}{\parsec}
\DeclareSIUnit{\kpc}{\kilo \parsec}
\DeclareSIUnit{\angstrom}{\AA}
\DeclareSIUnit{\Msun}{\ensuremath{\mathrm{M}_{\odot}}}
\DeclareSIUnit{\Lsun}{\ensuremath{\mathrm{L}_{\odot}}}
\DeclareSIUnit{\mag}{\ensuremath{\mathrm{mag}}}
\DeclareSIUnit{\year}{\ensuremath{\mathrm{yr}}}

\usepackage{hyperref} 
\hypersetup{
	unicode=true,
	colorlinks=true,
	linkcolor=blue,
	citecolor=blue,
	filecolor=magenta,
	urlcolor=blue
}

\makeatletter
\g@addto@macro{\UrlBreaks}{\UrlOrds}
\makeatother

\begin{document} 

   \title{ARRAKIHS:
   The observability of tidally stripped satellites in a Milky Way-like halo
   }

   \author{
    A.~Pérez-Herrero\inst{\ref{aff:ifca}}\fnmsep
    \thanks{Corresponding authors: Abram Pérez-Herrero, Yves Revaz, Santi Roca-Fàbrega, and Srikanth T. Nagesh.}
    \and
    Y. Revaz\inst{\ref{aff:epfl}}
    \and
    S. Roca-Fàbrega\inst{\ref{aff:lund}}
    \and
    S. T. Nagesh\inst{\ref{aff:epfl}}
    \and
    J. Preston\inst{\ref{aff:surrey}}
    \and
    D. Erkal\inst{\ref{aff:surrey}}
    \and
    D. Herranz Muñoz\inst{\ref{aff:ifca}}
    \and
    O. Agertz\inst{\ref{aff:lund}}
    \and
    M. Akhlaghi\inst{\ref{aff:cefca}}
    \and
    R. Bieri\inst{\ref{aff:zurich}}
    \and
    A. Camaz\'on-Pinilla\inst{\ref{aff:ifca},\ref{aff:florida}}
    \and
    J. Cenarro\inst{\ref{aff:cefca}}
    \and
    A. P. Cooper\inst{\ref{aff:nthu},\ref{aff:cica}}
    \and
    H. Domínguez Sánchez\inst{\ref{aff:ifca}}
    \and
    A. Fattahi\inst{\ref{aff:stockholm}}
    \and
    R. Feldmann\inst{\ref{aff:zurich}}
    \and
    A. Font\inst{\ref{aff:ljmu}}
    \and
    M. A. Gómez-Flechoso\inst{\ref{aff:ucm},\ref{aff:iparcos}}
    \and
    E. Gaztañaga\inst{\ref{aff:portsmouth},\ref{aff:ice},\ref{aff:ieec}}
    \and
    M. G\"udel\inst{\ref{aff:vienna}}
    \and
    R. Guzmán\inst{\ref{aff:ifca},\ref{aff:florida}}
    \and
    E. Illés\inst{\ref{aff:groningen},\ref{aff:tartu}}
    \and
    R. Infante-Sainz\inst{\ref{aff:cefca}}
    \and
    C. Kintziger\inst{\ref{aff:ghost}}
    \and
    A. Mar\'in-Franch\inst{\ref{aff:cefca}}
    \and    
    D. Martínez-Delgado\inst{\ref{aff:cefca},\ref{aff:araid}}
    \and
    B. Moore\inst{\ref{aff:zurich}}
    \and
    K. Nagamine\inst{\ref{aff:osaka}}
    \and
    P. Papaderos\inst{\ref{aff:papaderos}}
    \and
    P. Renard Guiral\inst{\ref{aff:ice},\ref{aff:ieec}}
    \and
    A. H.~Riley\inst{\ref{aff:lund}}
    \and
    R. Rodríguez-Cardoso\inst{\ref{aff:ucm},\ref{aff:iparcos}}
    \and
    S. Serrano\inst{\ref{aff:ice},\ref{aff:ieec},\ref{aff:Satlantis}}
    \and
    S. Shen\inst{\ref{aff:shen}}
    \and
    E. Sola\inst{\ref{aff:cambridge}}
    \and
    F. Torradeflot\inst{\ref{aff:ciemat},\ref{aff:PIC}}
    \and
    M. Wiesmann\inst{\ref{aff:shen}}
    }
    
    \institute{
        Instituto de Física de Cantabria (IFCA, CSIC-UC), Avenida de los Castros s/n, 39005 Santander, Spain
        \label{aff:ifca}
        \and
        Institute of Physics, Laboratory of Astrophysics, École Polytechnique Fédérale de Lausanne (EPFL), 1290 Sauverny, Switzerland
        \label{aff:epfl}
        \and
        Lund Observatory, Division of Astrophysics, Department of Physics, Lund University, SE-221 00 Lund, Sweden
        \label{aff:lund}
        \and
        School of Mathematics and Physics, University of Surrey, Guildford, UK
        \label{aff:surrey}
        \and
        Centro de Estudios de F\'isica del Cosmos de Arag\'on (CEFCA), Plaza San Juan, 1, E-44001, Teruel, Spain
        \label{aff:cefca}
        \and
        Universität Zürich, Institut für Astrophysik, Winterthurerstrasse 190, 8057 Zürich, Switzerland
        \label{aff:zurich}
        \and
        Department of Astronomy, University of Florida, 211 Bryant Space Sciences Center, Gainesville, FL 32611, USA
        \label{aff:florida}
        \and
        Institute of Astronomy and Department of Physics, National Tsing Hua University, Hsinchu 30013, Taiwan
        \label{aff:nthu}
        \and
        Center for Informatics and Computation in Astronomy, National Tsing Hua University, Hsinchu 30013, Taiwan
        \label{aff:cica}
        \and
        The Oskar Klein Centre, Department of Physics, Stockholm University, AlbaNova University Center, 106 91 Stockholm, Sweden
        \label{aff:stockholm}
        \and
        Astrophysics Research Institute, Liverpool John Moores University, 146 Brownlow Hill, Liverpool L3 5RF, UK
        \label{aff:ljmu}
        \and
        Departamento de Física de la Tierra y Astrofísica, Fac. CC. Físicas, Universidad Complutense de Madrid, Plaza Ciencias, 1, 28040 Madrid, Spain
        \label{aff:ucm}
        \and
        Instituto de Física de Partículas y del Cosmos (IPARCOS), Fac. CC. Físicas, Universidad Complutense de Madrid, Plaza Ciencias, 1, 28040 Madrid, Spain
        \label{aff:iparcos}
        \and
        Institute of Cosmology \& Gravitation, University of Portsmouth, Portsmouth PO1 3FX, United Kingdom
        \label{aff:portsmouth}
        \and
        Institute of Space Sciences (ICE, CSIC), Campus UAB, Carrer de Can Magrans, s/n, 08193 Barcelona, Spain
        \label{aff:ice}
        \and
        Institut d'Estudis Espacials de Catalunya (IEEC), 08034 Barcelona, Spain
        \label{aff:ieec}
        \and
        Department of Astrophysics, University of Vienna,
        Türkenschanzstrasse 17, A-1180 Vienna, Austria
        \label{aff:vienna}
        \and
        Kapteyn Astronomical Institute, University of Groningen, Landleven 12, NL-9747 AD Groningen, the Netherlands
        \label{aff:groningen}
        \and
        Tartu Observatory, University of Tartu, Observatooriumi 1, 61602 Tõravere, Estonia
        \label{aff:tartu}
        \and
        GHOST Space, Belgium \label{aff:ghost}
        \and
        ARAID Foundation, Avda. de Ranillas, 1-D, E-50018 Zaragoza, Spain
        \label{aff:araid}
        \and
        Theoretical Astrophysics, Department of Earth and Space Science, The University of Osaka, 1-1 Machikaneyama, Toyonaka, Osaka 560-0043, Japan
        \label{aff:osaka}
        \and
        Instituto de Astrofísica e Ciências do Espaço, Universidade do Porto – CAUP, Rua das Estrelas, PT4150-762 Porto, Portugal\label{aff:papaderos}
        \and
        Satlantis, University Science Park, Sede Bld 48940, Leioa-Bilbao, Spain
        \label{aff:Satlantis}
        \and
        Institute of Theoretical Astrophysics, University of Oslo, P.O. Box 1029 Blindern, N-0315 Oslo, Norway.
        \label{aff:shen}
        \and
        Institute of Astronomy, University of Cambridge, Madingley Road, Cambridge CB3 0HA, UK
        \label{aff:cambridge}
        \and
        Centro de Investigaciones Energéticas, Medioambientales y
        Tecnológicas (CIEMAT), Avenida Complutense 40,
        28040 Madrid, Spain \label{aff:ciemat}
        \and
        Port d'Informació Científica (PIC), Campus UAB,
        C. Albareda s/n, 08193 Bellaterra (Barcelona), Spain \label{aff:PIC}
    }
   \date{\today}
 
    \abstract
{Current and forthcoming deep imaging surveys are revealing an increasing number of low surface brightness stellar streams around Milky Way-like galaxies. These streams encode information about the orbital evolution and internal structure of their disrupted dwarf galaxy progenitors. We investigate the dynamical evolution and observability of tidal features produced by dwarf galaxies consistent with the cosmological constant and cold dark matter (\(\Lambda\)CDM) framework and observations of Local Group satellites. We performed high-resolution $N$-body simulations of dwarf galaxies orbiting within a static Milky Way-like potential comprising a disc, a bulge, and a dark matter halo. Each dwarf comprised two collisionless components: a Plummer stellar distribution embedded in a cuspy or cored generalised Navarro--Frenk--White dark matter halo. Idealised integrated surface brightness maps were constructed from simulation outputs every \(0.5\,\mathrm{Gyr}\) over \(10\,\mathrm{Gyr}\), and detectability was quantified through the projected area of progenitor stellar light brighter than a range of limiting surface brightnesses, \(\mu_{\rm lim}=29\)--\(32\,\mathrm{mag\,arcsec^{-2}}\), 
chosen to span depths relevant to the Analysis of Resolved Remnants of Accreted galaxies as a Key Instrument for Halo Surveys (ARRAKIHS), the Legacy Survey of Space and Time (LSST), and \textit{Euclid}. The detectability of tidal features depends most strongly on progenitor stellar mass, orbital configuration, parameterised by the eccentricity proxy \(e_0\), and evolutionary stage. Over the surface brightness range explored, extended tidal debris is predominantly recovered for initially more radial configurations, mainly those with \(e_0 \gtrsim 0.75\). In the simulated grid, detectable area is dominated by \(M_\star=10^8\)--\(10^9\,M_\odot\) progenitors, whereas \(M_\star=10^6\)--\(10^7\,M_\odot\) systems can undergo stripping without detectable debris. The detectable area evolves non-monotonically with time, reflecting the competition between stream formation, phase mixing, and the fading of diffuse debris below the adopted surface brightness limits. Cuspy and cored progenitors show similar evolution of the detectable area, with differences generally smaller than the scatter induced by orbital diversity.}
    
   \titlerunning{ARRAKIHS: The observability of tidally stripped satellites}
   \authorrunning{A. Pérez-Herrero et al.}
   \maketitle
\section{Introduction}
Stellar tidal streams, shells, and other diffuse structures have already been observed around nearby galaxies through both targeted deep imaging and wide-field surveys \citep[e.g.][]{2010AJ....140..962M,2016MNRAS.456.1359F}. These observations have enabled the identification and characterisation of a wide variety of tidal features. Despite this progress, the recovery of the faintest and most diffuse tidal structures remains observationally challenging. Current and forthcoming generations of deep, wide-field imaging surveys will extend these studies by probing fainter structures and providing larger and more systematic samples. Facilities such as \textit{Euclid}, the \textit{Nancy Grace Roman} Space Telescope, and the \textit{Vera C. Rubin} Observatory Legacy Survey of Space and Time (LSST) will improve the detection of diffuse stellar structures in the outskirts of external galaxies \citep[e.g.][]{Euclid2022,RomanLSB,lsst}. Dedicated low surface brightness (LSB) surveys and missions, such as the Large Binocular Telescope Imaging of Galactic Halos and Tidal Structures survey (LIGHTS) and the adopted mission Analysis of Resolved Remnants of Accreted galaxies as a Key Instrument for Halo Surveys (ARRAKIHS), are specifically designed to explore this faint regime \citep[e.g.][]{LIGTH}. Complementary surveys, such as the Stellar Stream Legacy Survey, exploit imaging from the Dark Energy Spectroscopic Instrument (DESI) Legacy Imaging Surveys to identify stellar streams around nearby galaxies \citep{Delgado}. These observations are expected to further expand the known population of faint halo substructures, including streams, shells, and diffuse debris, whose properties provide sensitive probes of the tidal evolution of satellite galaxies \citep[e.g.][]{Vera_Casanova_2022,Mir_Carretero_2024,stoiber2025}.

As dwarf galaxies orbit within the gravitational potential of their host haloes, tidal forces remove stars from their outskirts and redistribute them into extended stellar structures. Previous studies have shown that the morphology and spatial distribution of tidal debris are shaped by the satellite's orbital history, progenitor luminosity, and viewing geometry \citep[e.g.][]{Johnston_2008,Hendel_Johnston_2015}. These properties are also affected by the satellite's internal structure, including the inner density profile of its dark matter halo, as well as by the global properties of the host potential \citep[e.g.][]{Erkal_2016,Errani_2019,Errani_2022}. The inner dark matter profile is particularly relevant in dwarf galaxies because of the long-standing cusp--core problem and because the central dark matter slope can affect the resilience of satellites to tidal stripping \citep[e.g.][]{de_Blok_2009,Penarubia_2010,Pontzen_2012}. These dependences motivate a controlled exploration of how orbital configuration, progenitor structure, and observational depth jointly affect the morphology and detectability of tidal debris.

To address this problem, simulations are required to investigate the disruption of dwarf galaxies. Among simulation-based approaches, cosmological simulations provide the most complete framework for studying the evolution of satellite galaxies and the debris produced by tidal stripping. High-resolution hydrodynamical zoom-in simulations and cosmological volumes, including Auriga \citep[e.g.][]{Grand_2017,Simpson_2018,Riley_2025,Shipp_2025}, Feedback In Realistic Environments (FIRE) simulations \citep[e.g.][]{Panithanpaisal_2021,Shipp_2023}, FIREbox \citep{2023MNRAS}, and VINTERGATAN \citep{Agertz_2021}, follow galaxy assembly in a fully cosmological context, capturing baryonic physics, time-dependent host potentials, and complex merger histories. In these environments, tidal stripping arises self-consistently as satellites orbit their hosts, producing stellar streams, partially disrupted systems, and phase-mixed structures \citep[e.g.][]{Panithanpaisal_2021, Riley_2025}. These studies show that disrupting satellites are common around Milky Way-like haloes and occupy a broad range of orbital and evolutionary states. However, the complexity and computational cost of cosmological simulations make it difficult to isolate the different factors that shape satellite evolution. Beyond the properties of the progenitor and its orbit, satellites are influenced by baryonic components, preprocessing in group environments, interactions with massive companions, and halo-to-halo variations in accretion history \citep[e.g.][]{Simpson_2018,Shipp_2025}. Even when using zoom-in techniques, evolving hydrodynamical simulations to \(z=0\) while maintaining sufficiently high resolution across the relevant dwarf galaxy mass range remains computationally demanding. In addition, many current high-resolution hydrodynamical simulations do not necessarily resolve the internal structure of low-mass dwarf galaxies and their tidal debris at the level required to trace the detailed disruption process, particularly in the mass regime that may contribute substantially to the stream population \citep{Shipp_2023}. Predictions from cosmological simulations can also be affected by the adopted subgrid prescriptions for unresolved baryonic physics, including star formation, feedback, and the effective equation of state used to model the unresolved interstellar medium \citep[e.g.][]{Shipp_2023,Pakmor_2025}. These limitations can modify the internal structure of dwarf galaxies and their response to tides, for example, through numerical heating, overly extended stellar components, or enhanced susceptibility to disruption \citep{Cooper_2017}. More generally, cosmological simulations may suffer from numerical over-merging, namely the artificial disruption of dark matter subhaloes due to insufficient resolution \citep{vandenbosch2018a,vandenbosch2018b,Errani_2019,benson2022,santossantos2025}. Disentangling the physical drivers of disruption from these numerical and modelling effects, and connecting them to the observability of the stripped stellar material, remains challenging.

Alternative semi-analytic approaches have also been developed to explore the disruption of satellite galaxies and the range of tidal debris morphologies produced by dwarf galaxy progenitors across wide parameter spaces \citep[e.g.][]{StreamGen}. These methods provide an efficient way to investigate trends in orbital evolution and mass loss in the satellite galaxy regime, but they rely on simplified prescriptions and do not follow the full gravitational evolution of self-consistent multi-component progenitors.

Idealised $N$-body simulations of realistic dwarf galaxies infalling into a Milky Way-mass potential have previously been employed to investigate several aspects of tidal stripping. Unlike fully cosmological simulations, these studies follow satellite evolution outside a cosmological context, using controlled progenitor models embedded in host galaxies. This reduction in complexity makes it possible to reach substantially higher resolution in the satellite and its tidal debris, allowing the internal structure of the progenitor and the details of the stripping process to be traced more accurately. It also makes it possible to isolate the role of progenitor structure and orbital configuration, including the structural evolution of dark matter haloes and the dependence of mass loss on the inner density profile and orbital parameters \citep[e.g.][]{Errani_2019, Aguirre_2025, arias_2025}. Such studies have provided important insights into the dynamical response of satellite galaxies to tidal forces in controlled environments. Nevertheless, the connection between progenitor structure and the surface brightness evolution of the stripped stellar component has not yet been explored systematically, limiting quantitative predictions for future LSB observations.

In this work, we constructed equilibrium two-component dwarf galaxy models with properties consistent with both the cosmological constant and cold dark matter (\(\Lambda\)CDM) framework and observations of Local Group satellites \citep{McConnachie_2020}, as well as with the size--luminosity relation reported by \citet{elves}. We evolved these systems in a controlled suite of high-resolution $N$-body simulations within a static Milky Way-like potential including disc, bulge, and halo components, together with an analytic dynamical friction prescription. With this setup, we quantified the dependence of tidal feature detectability on progenitor stellar mass, inner dark matter density profile, orbital configuration, evolutionary stage, and line of sight. From the simulation snapshots, we used \texttt{pNbody} \citep{revaz_pnbody_2013} to generate idealised integrated surface brightness maps in the ARRAKIHS VIS2 band and measure the evolution of feature detectability, thereby establishing trends that link progenitor structure to observable LSB signatures. This controlled framework allows us to identify the physical conditions under which stellar streams are expected to be detectable in current and forthcoming deep imaging surveys and to assess whether changes in the inner dark matter density profile leave observable signatures in the properties of the streams.

The remainder of this paper is organised as follows. Section~\ref{sec:methods_simulations} describes the dwarf galaxy progenitor models, host potential and dynamical-friction prescription, the orbital setup, and the methodology used to generate mock surface brightness maps and quantify feature detectability. Section~\ref{sec:results} presents the dependence of detectability on orbital configuration, evolutionary stage, progenitor stellar mass, and inner dark matter density profile. Section~\ref{sec:caveats} discusses the main caveats of our idealised setup and outlines future directions. Section~\ref{sec:conclusion} summarises our main conclusions. 

\section{Methods and simulations}\label{sec:methods_simulations}

In this section, we describe the construction of the dwarf galaxy progenitor models, the numerical setup of the simulations, and the methodology used to generate mock surface brightness maps and quantify debris detectability.

\subsection{Dwarf galaxy progenitor models}\label{sec:progenitors}

The progenitor dwarf galaxies were modelled as two-component systems in dynamical equilibrium composed of a stellar distribution embedded within a dark matter halo. The stellar component follows a Plummer density profile \citep{Plummer1911},
\begin{equation}
\rho_\star(r) =
\frac{3M_\star}{4\pi a^3}
\left(1 + \frac{r^2}{a^2}\right)^{-5/2},
\end{equation} 
where $M_\star$ is the total stellar mass, and $a$ is the Plummer scale radius.

The dark matter halo was described by a generalised Navarro--Frenk--White density profile \citep[e.g.][]{2001ApJ...555..504W}, corresponding to a particular case of the general double power-law family introduced by \citet{Zhao_1996},
\begin{equation}
\rho_{\rm DM}(r) =
\frac{\rho_0}
{(r/r_s)^\gamma (1+r/r_s)^{3-\gamma}},
\end{equation}
where $\gamma$ controls the inner logarithmic slope and the outer slope is fixed to $-3$. This corresponds to the Zhao profile with its transition parameter fixed to unity and $\beta=3$. We considered both cuspy ($\gamma=1$) and cored ($\gamma=0$) configurations. These two configurations bracket different possible central dark matter structures and allow us to assess how the inner slope affects the tidal evolution and observability of the stellar component.

The structural parameters were chosen to be consistent with Local Group dwarf galaxies over the stellar-mass range \(M_\star = 10^{6}\)--\(10^{9}\,M_\odot\) \citep{McConnachie_2020, elves}. The adopted parameters are summarised in Table~\ref{tab:table_models}. For each stellar-mass scale, both cuspy and cored configurations were explored.

\begin{table*}[t]
\centering
\caption{Structural parameters of the dwarf galaxy progenitor models.}
\label{tab:table_models}

\begin{tabular}{r r r r r r r}
\toprule
{$M_{200}$ [$M_\odot$]} & {$c_{200}$} & {$\gamma$} &
{$M_\star$ [$M_\odot$]} & {$a$ [kpc]} &
{$\langle$[Fe/H]$\rangle$} & {$\sigma_{\rm los}$ [km s$^{-1}$]} \\
\midrule
$1\times10^{11}$ & 10 & 1 & $10^{9}$ & 2.0 & -0.5 & 40.0 \\
$1\times10^{11}$ & 30 & 0 & $10^{9}$ & 2.0 & -0.5 & 38.4 \\
$2\times10^{10}$ & 11 & 1 & $10^{8}$ & 1.1 & -1.1 & 25.8 \\
$2\times10^{10}$ & 30 & 0 & $10^{8}$ & 1.1 & -1.1 & 23.2 \\
$2\times10^{9}$  & 12 & 1 & $10^{7}$ & 0.6 & -1.5 & 13.3 \\
$2\times10^{9}$  & 29 & 0 & $10^{7}$ & 0.6 & -1.5 & 12.2 \\
$4\times10^{8}$  & 16 & 1 & $10^{6}$ & 0.3 & -1.9 & 7.9 \\
$4\times10^{8}$  & 36 & 0 & $10^{6}$ & 0.3 & -1.9 & 7.8 \\
\bottomrule
\end{tabular}

\tablefoot{
The concentration is defined as $c_{200}=r_{200}/r_s$.
The parameter $\gamma$ denotes the inner logarithmic
slope of the dark matter density profile; the outer
slope is fixed to $\beta=3$ for all models.
}

\end{table*}

For the cuspy models, halo concentrations were assigned using the mass--concentration relation evaluated at redshift \(z=0\) \citep{Dutton:2014xda}. This choice does not imply that the progenitors are assumed to undergo late cosmological growth during our simulations. Rather, the initial conditions are equilibrium models intended to represent already virialised dwarf galaxy haloes. Low-mass haloes are expected to collapse and virialise at high redshift \citep{revaz2009}; after this stage, they become gravitationally bound systems whose physical inner density profiles do not evolve significantly in the absence of strong external perturbations. In contrast, a substantial part of the apparent evolution of virial quantities such as \(M_{200}\), \(R_{200}\), and concentration arises from pseudo-evolution, that is, from the changing reference density used to define the virial radius, rather than from a genuine restructuring of the inner halo \citep{Diemer_2013}. Since our models were calibrated to match the structural and kinematic properties of present-day Local Group dwarf galaxies, adopting the \(z=0\) mass--concentration relation provides a suitable description of their physical halo structure.

For the cored models, the halo concentration was adjusted so that the stellar velocity dispersion remained consistent with the observed Local Group scaling relations and comparable to that of the corresponding cuspy model. We adjusted the concentration rather than \(M_{200}\) because the stellar velocity dispersion is primarily set by the mass enclosed within the stellar body, which is more directly controlled by the halo concentration than by the total virial mass. We therefore did not keep the halo concentration fixed when constructing the cusp--core pairs. At fixed concentration, replacing a cusp by a core would lower the central gravitational potential and would produce stellar velocity dispersions that are too small compared with the observed dwarf galaxy population. Our comparison should thus be interpreted as a comparison between observationally plausible cuspy and cored progenitors with the same stellar component and similar stellar kinematics, rather than as an isolated variation of the inner logarithmic slope at fixed global halo structure.

\begin{figure}
    \centering
    \includegraphics[width=\linewidth]{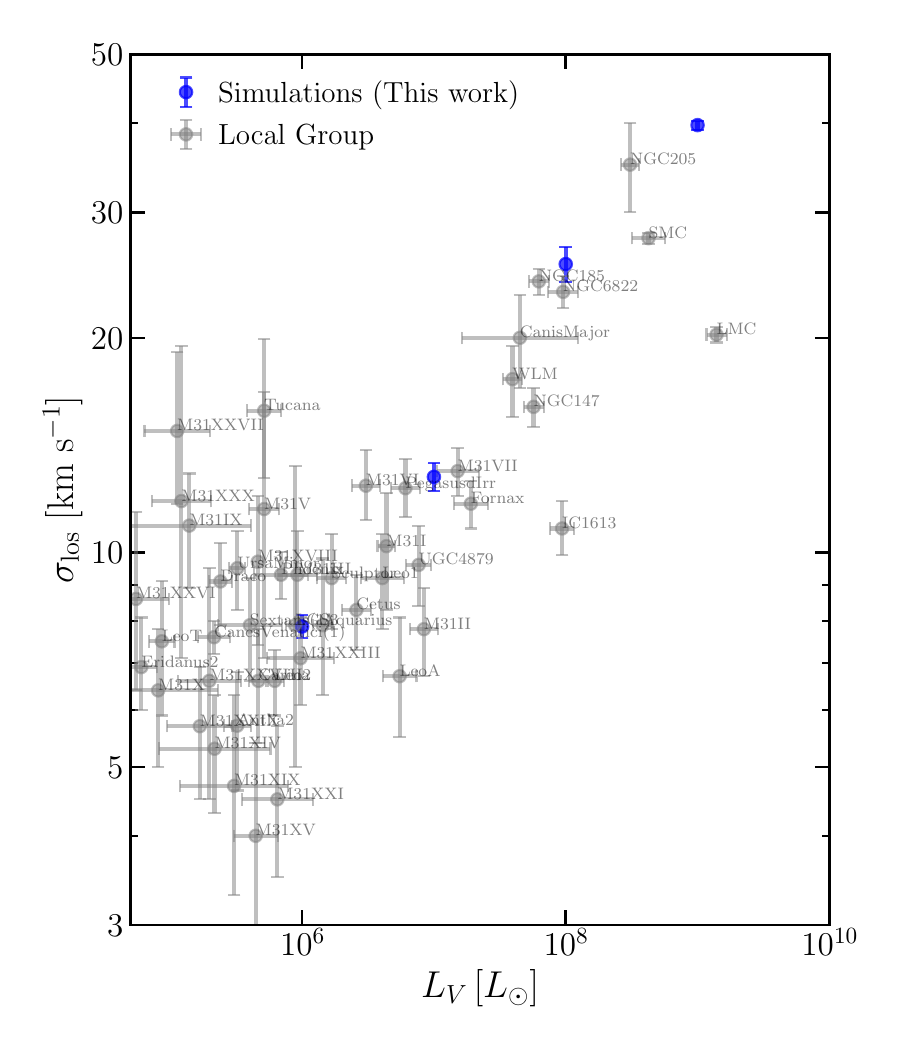}
    \caption{Stellar line-of-sight velocity dispersion measured at 1 kpc as a function of V-band luminosity. Grey symbols show Local Group dwarf galaxies from \citet{McConnachie_2020}, while blue symbols correspond to the simulated galaxies presented in this work. Error bars indicate the dispersion measured along different lines of sight.}
    \label{fig:L_sigma}
\end{figure}

\begin{figure}
    \centering
    \includegraphics[width=\linewidth]{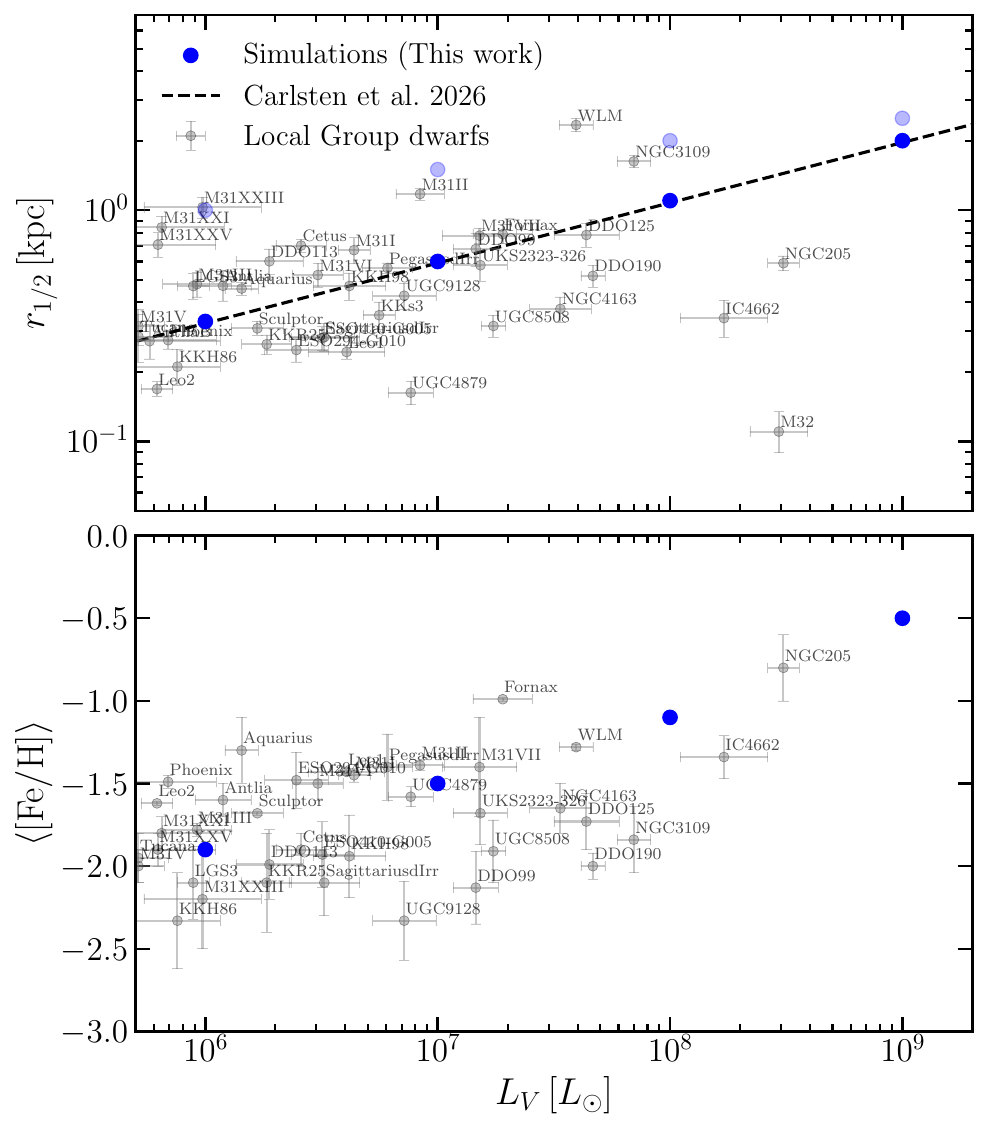}
    \caption{Top panel: Projected half-light radius $r_{1/2}$ as a function of V-band luminosity $L_V$. The dashed line shows the best-fitting relation reported by \citet{elves}. Bottom panel: Mean stellar metallicity $\langle \mathrm{[Fe/H]} \rangle$ as a function of $L_V$. Grey symbols indicate Local Group dwarf galaxies \citep{McConnachie_2020}. Blue symbols show our simulated galaxies, while lighter blue symbols correspond to our more extended simulations discussed in Appendix~\ref{apx:additional_models}.
    }
\label{fig:Rh_FeH_vs_L}
\end{figure}

Figures~\ref{fig:L_sigma} and~\ref{fig:Rh_FeH_vs_L} compare the structural and kinematic properties of our progenitor models with observed dwarf galaxy scaling relations. Figure~\ref{fig:L_sigma} shows the line of sight averaged stellar velocity dispersion, \(\langle\sigma_{\rm los}\rangle\), measured at a projected radius of \(1\,\mathrm{kpc}\), as a function of \(V\)-band luminosity. Figure~\ref{fig:Rh_FeH_vs_L} shows the projected half-light radius and mean stellar metallicity as a function of luminosity. To compute \(\langle\sigma_{\rm los}\rangle\), we rotated each equilibrium model along 100 randomly oriented viewing directions. For each viewing direction, which defines a line of sight, we projected the stellar velocities onto that direction and measured the dispersion of the line of sight velocity component within the projected radial bin containing \(1\,\mathrm{kpc}\). The values shown in Fig.~\ref{fig:L_sigma} correspond to the mean over the 100 projections, with the error bars indicating the 1-\(\sigma\) scatter among viewing directions. This averaging reduces the sensitivity of the measurement to a particular projection. The comparison was made with the observed Local Group and nearby dwarf galaxy compilation of \citet{McConnachie_2020}, as well as with the size--luminosity relation reported by \citet{elves}. We used these data as a reference for the structural and kinematic properties of dwarf galaxies. The models lie within the observed scatter, indicating that they provide plausible initial equilibrium models for dwarf galaxy progenitors. 

The mean metallicity shown in Fig.~\ref{fig:Rh_FeH_vs_L} was used to define the metallicity parameter of each progenitor. Individual stellar particles were then assigned metallicities from an asymmetric distribution with a metal-poor tail, obtained by logarithmically transforming values drawn from an exponential distribution. Stellar ages were assigned separately, using mass-dependent parametric distributions calibrated against the mean star-formation histories of the chemo-dynamical dwarf galaxy models of \citet{revaz_2018}. To impose an age--metallicity relation in which metal-poor stars are the oldest, we assigned the oldest stellar particles to the most metal-poor metallicity values. Since this procedure would otherwise produce an artificially tight age--metallicity relation, we supplemented the metallicities with a Gaussian intrinsic scatter. In particular, low mass dwarfs were represented by predominantly old stellar populations, corresponding to systems with early-quenched star formation, whereas more massive progenitors were allowed to have more extended star-formation histories. No radial age gradient was imposed in the initial stellar component.

The main limitation of this approximation concerns the stellar component. Since the stellar masses and luminosities adopted here were calibrated to present-day dwarfs, they may overestimate the stellar mass already in place at early infall times, particularly for more massive or continuously star-forming systems. Conversely, the absence of very young stellar populations in the most massive progenitors could underestimate their instantaneous luminosity. This effect may be more important for massive progenitors, which can continue forming stars for several gigayears after infall \citep[e.g.][]{Akins_2021,Rodr_guez_Cardoso_2025}. In such cases, however, residual star formation is expected to be concentrated mainly in the inner regions, since the outer cold gas reservoir is preferentially stripped. For lower-mass systems, or after extensive mass loss, star formation is expected to be rapidly suppressed, and the luminosity contribution from very young populations decreases quickly after \(\sim10\,\mathrm{Myr}\). We therefore do not expect this simplification to significantly affect the relative trends in debris detectability over the gigayear timescales considered here.

Initial conditions were generated using \texttt{pNbody} \citep{revaz_pnbody_2013}, sampling independently the distribution functions of the stellar and dark matter components self-consistently within the total gravitational potential. The sampling procedure followed the methodology described in \citet{Errani_2019}, ensuring that the composite systems are in dynamical equilibrium.

Each progenitor model was sampled with a minimum of \(10^4\) stellar particles to adequately resolve the internal structure of the dwarf galaxies and mitigate artificial two-body relaxation. This corresponds to stellar particle masses of \(m_\star = 10^2\), \(10^3\), and \(10^4\,\mathrm{M_\odot}\) for \(M_\star = 10^6\), \(10^7\), and \(10^8\)--\(10^9\,\mathrm{M_\odot}\) models, respectively. The dark matter particles were assigned the same mass as the stellar particles in the corresponding progenitor model, with \(m_{\rm DM}=10^2\), \(10^3\), and \(10^4\,\mathrm{M_\odot}\) for the same three stellar-mass ranges.

We tested the impact of numerical resolution by repeating a representative subset of simulations at particle numbers one order of magnitude higher than in the fiducial runs. The resulting evolution of the detectable area, \(A_{\rm det}\), defined in Sect.~\ref{sec:sb_maps}, remains stable under this change in resolution, with no systematic shift in the measured values. The main trends with stellar mass, orbital eccentricity, and limiting surface brightness are also not driven by the adopted particle sampling.

\begin{figure}
    \centering
    \includegraphics[width=\linewidth]{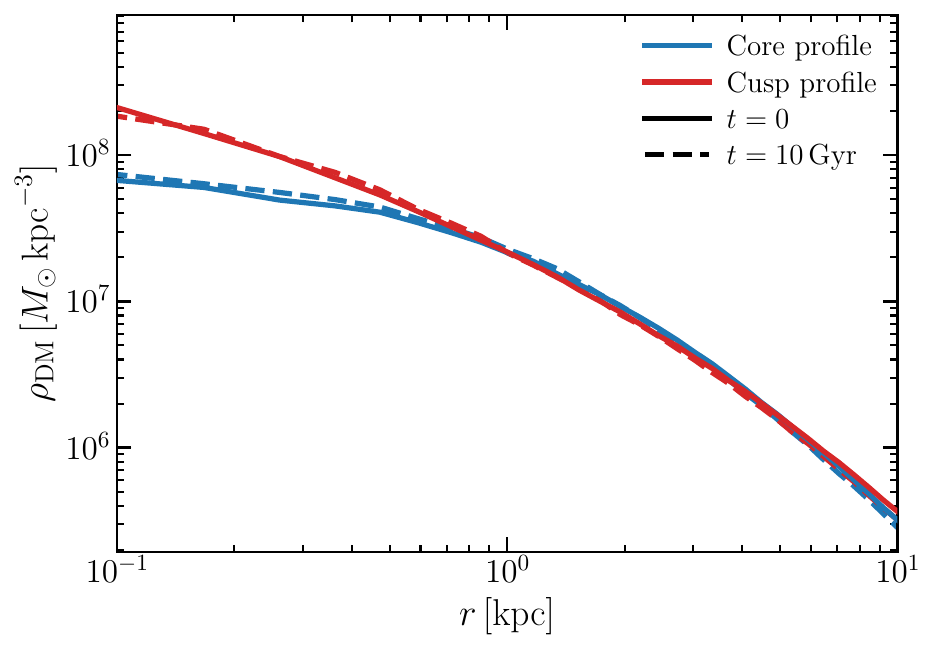}
    \caption{Dark matter density profile as a function of radius for representative cuspy (red) and cored (blue) models. Solid lines show the initial profiles, while dashed lines indicate the profiles after 10\,Gyr of isolated evolution.}
    \label{fig:Density_profile}
\end{figure}

Figure~\ref{fig:Density_profile} shows the isolated evolution of the dark matter density profile for representative cuspy and cored models. In both cases, the density profiles remain stable, and cuspy haloes preserve their inner logarithmic slope, showing that numerical relaxation does not induce artificial core formation.

\subsection{Simulation setup and host potential}\label{sec:simulation_framework}

All simulations presented in this work were performed with the public $N$-body code \textsc{SWIFT} (SPH With Inter-dependent Fine-grained Tasking) \citep{swift_code}\footnote{The runs presented here used \textsc{SWIFT} version 2025.04.}, using a collisionless setup in which the dwarf galaxy progenitors evolved under their self-gravity while orbiting within an externally imposed Milky Way-like potential, as described below.

Self-gravity was computed with the fast multipole method implemented in \textsc{SWIFT}, using an adaptive multipole-acceptance criterion. Particle time-steps followed the standard acceleration-based criterion with \(\eta=0.025\), where \(\eta\) sets the time-step accuracy parameter. The simulations did not include gas physics, star formation, or feedback processes.

The maximum gravitational softening was set to $\epsilon = 20\,\mathrm{pc}$ for both dark matter and stellar particles. This value is more than an order of magnitude smaller than the smallest stellar scale radius considered, ensuring that the internal structure of the progenitors is well resolved. Each model was evolved for a total integration time of 10\,Gyr.

The host galaxy was represented by a static background potential in \textsc{SWIFT} (Roduit et al., in prep.), based on the Milky Way model of \citet{bovy2015}. This potential consists of a Navarro--Frenk--White halo \citep{Navarro_1997}, a Miyamoto--Nagai disc \citep{Miyamoto_Nagai_1975}, and a spherical bulge with an exponential cut-off.

Since the host potential is analytic, orbital decay was included by adding an analytic Chandrasekhar dynamical friction term acting on the satellite centre of mass. The dynamical-friction acceleration was computed using the initial satellite mass $M_{\rm sat}$ and is written following Chap.~8, Eq.~8.7 in \citet{binney2008}:
\begin{equation}
\frac{\mathrm{d}\vec{v}}{\mathrm{d}t}
=
-\frac{4\pi G^2 M_{\rm sat}\,\rho(r)\,\ln\Lambda}{v^3}
\left[
\mathrm{erf}(X) - \frac{2X}{\sqrt{\pi}}e^{-X^2}
\right]\vec{v},
\end{equation}
where \(G\) is the gravitational constant, \(M_{\rm sat}\) is the initial satellite mass, \(\vec{v}\) is the velocity of the satellite centre of mass relative to the host, \(v=|\vec{v}|\), \(r\) is the galactocentric distance of the satellite, \(\ln\Lambda\) is the Coulomb logarithm, \(X=v/[\sqrt{2}\sigma(r)]\), and \(\mathrm{erf}\) is the error function. The quantities \(\rho(r)\) and \(\sigma(r)\) are the spherically averaged density and one-dimensional velocity dispersion of the host halo, respectively, computed from the adopted host-halo model.

The Coulomb logarithm was fixed to \(\ln\Lambda=1.0\), based on a calibration against a representative subset of otherwise identical simulations evolved in a live Milky Way halo. In these calibration runs, we compared the time evolution of the satellite galactocentric radius, the sequence of pericentric passages, and the cumulative orbital decay over 10 Gyr. This value provides the closest match to the live-halo trajectories for the satellite masses considered here.

This prescription is approximate. Since the friction term was evaluated using the initial satellite mass, it can mildly overestimate the drag once a satellite has lost a significant fraction of its mass. This effect is expected to be most relevant for the most massive progenitors and for systems undergoing strong disruption \citep{Trelles_2022}. We therefore interpret the detailed late-time orbital evolution of these systems as approximate, while using the calibrated prescription to capture the global orbital-decay trends across the simulated grid. This caveat is considered when interpreting the evolved orbital properties in Sect.~\ref{sec:results_orbit}.

To avoid unrealistically strong drag in regions where the adopted \(\sigma(r)\) model becomes very small, we imposed a floor of \(\sigma_{\rm min}=10\,\mathrm{km\,s^{-1}}\) and smoothly suppressed the friction inside \(r_{\rm core}=10\,\mathrm{kpc}\), preventing excessive accelerations near the Galactic centre.

This framework provides a controlled and computationally efficient approximation to orbital decay, enabling a systematic exploration of progenitor structure and orbital parameters while maintaining high internal resolution in the satellite.

\subsection{Orbital setup and orbital parameters}\label{sec:orbits}

We considered eight progenitor dwarf galaxy models, corresponding to four stellar masses, each realised with both cuspy and cored dark matter haloes. Each of these eight progenitor models was evolved independently along the same set of 49 bound trajectories within the static Milky Way potential described in Sect.~\ref{sec:simulation_framework}, resulting in a total of 392 dynamical simulations. The initial coordinates and velocities were selected to span a broad range of galactocentric distances, eccentricities, and orbital orientations. Although this orbital sample is not intended to reproduce the full cosmological distribution of satellite orbits, it provides sufficient coverage to identify the main trends with orbital configuration in our controlled setup.

At injection, the galactocentric distances ranged from $\sim 80$ to $\sim 250\,\mathrm{kpc}$, allowing us to probe how tidal disruption depends on orbital distance. For each system, we computed the tidal radius in the circular orbit approximation at the initial galactocentric distance and used this value as the outer truncation radius of the initial satellite model. Both the stellar and dark matter components were sampled only within this radius, ensuring that no particles were initially placed outside the initial tidal radius.

The initial orbital parameters of the 49 trajectories selected for this work are provided in Table~\ref{tab:objects}. The corresponding values of \(e_0\) were used as a measure of initial orbital eccentricity. They were computed from the initial position and velocity of each satellite using the standard Keplerian expression for a point-mass host with \(M_{\rm host}=10^{12}\,M_\odot\). Since the subsequent evolution was followed in the full Milky Way potential, these values should be interpreted as an initial eccentricity proxy rather than as exact eccentricities of closed orbits. The explored configurations extend from highly radial to moderately eccentric orbits. Nearly circular cases were excluded because test simulations show that, at the adopted injection radii, they do not reach sufficiently small galactocentric distances, nor spend enough time in the strong tidal field of the host, to undergo significant stripping within the \(10\,\mathrm{Gyr}\) evolution. A variety of orbital inclinations was also considered to sample different geometries relative to the host disc.

The dynamical evolution was characterised by tracking the time-dependent position of the satellite density peak. At each snapshot, this position was estimated from the stellar distribution using a density-based centre, in which local stellar densities were computed from the nearest neighbours and the centre was associated with the highest-density region \citep{casertano}. This procedure provides a robust estimate of the satellite centre while a bound stellar remnant is still identifiable. The sequence of measured centres was then complemented by orbit integrations in the same Milky Way potential adopted for the simulations, which provide a continuous radial trajectory and reduce the sensitivity of the inferred pericentric and apocentric passages to the discrete snapshot spacing. Pericentres and apocentres were identified as successive local minima and maxima of this continuous radial trajectory. At late times, when the progenitor is strongly disrupted and the stellar distribution no longer defines a unique centre, the integrated orbit was used only as a reference trajectory for assigning the corresponding orbital phase. These evolved pericentre--apocentre pairs were used in Sect.~\ref{sec:results_orbit} to connect the orbital evolution of each progenitor with its morphological classification and detectable area.

\subsection{Mock surface brightness maps}\label{sec:sb_maps}

To quantify the detectability of progenitor stellar light as a function of progenitor mass, inner halo profile, orbital configuration, and evolutionary time, we constructed mock surface brightness maps of the stellar component throughout the full evolution of each system.

Simulation snapshots were analysed every $0.5\,\mathrm{Gyr}$. For each snapshot, we used \texttt{pNbody} to generate two-dimensional surface brightness mock images.
Each stellar particle was treated as a single stellar population with an assigned age and metallicity. We assigned a spectral energy distribution to each particle using the Binary Population and Spectral Synthesis model (BPASS, version 2.3; \citealt{byrne2022}). 

The mean flux \(\langle f_{\nu} \rangle\) in
\(\mathrm{erg\,s^{-1}\,cm^{-2}\,Hz^{-1}}\) in a given band is obtained after integrating over the filter response $R(\lambda)$ \citep[e.g.][]{willmer2018}:
\begin{equation}
\langle f_{\nu} \rangle = \frac{\int f(\lambda) R(\lambda) \lambda d\lambda}{\int R(\lambda)\frac{c}{\lambda} d\lambda},
  \label{eq:mean_flux}
\end{equation}
where $c$ is the speed of light and $f(\lambda)$ is the flux density per unit wavelength in 
\(\mathrm{erg\,s^{-1}\,cm^{-2}}\,\text{\AA}^{-1}\).
The magnitude in the AB system is then obtained using:

\begin{equation}
M_{\rm{AB}} =  -2.5 \,\log_{10}\left( \langle f_{\nu} \rangle \right) -48.6,
\end{equation} 
where the zero point (\(-48.6\)) ensures the magnitude to be precisely zero in any filter 
for a flux equal to $3631\,\rm{Jy}$\footnote{$3.631\cdot 10^{-20}\,\rm{erg\cdot s^{-1} \cdot cm^{-2} 
\cdot Hz^{-1}}$}.

We adopted the ARRAKIHS VIS2 band, covering approximately $400$--$900\,\mathrm{nm}$, as our fiducial configuration, which is similar to the \textit{Euclid} VIS band \citep{refId0}, and placed all systems at a distance of $35\,\mathrm{Mpc}$. This distance was chosen as a representative value for ARRAKIHS-like observations of LSB tidal features around nearby external galaxies. The AB absolute magnitudes of the stellar particles were converted into apparent magnitudes using the corresponding distance modulus, $m_i = M_i + 5\log_{10}\left[D/(10\,\mathrm{pc})\right]$, which gives $m_i=M_i+32.72$ for $D=35\,\mathrm{Mpc}$. Following the implementation in \texttt{pNbody}, these apparent magnitudes were converted into zero-point-normalised fluxes, $f_i = 10^{-m_i/2.5}$, which were then projected and summed on the image grid. The resulting flux map was converted into surface brightness using $\mu_j = -2.5\log_{10}(F_j/\Omega_{\rm pix})$, where $F_j$ is the total zero-point-normalised flux in pixel $j$ and $\Omega_{\rm pix}$ is the angular area of the image pixel. For the adopted VIS2 configuration, the pixel scale is $1.375\,\mathrm{arcsec}$ per pixel, corresponding to $\Omega_{\rm pix}=1.8909\,\mathrm{arcsec^2}$.

To reduce discreteness noise in the projected maps due to the discretisation of the stellar component into particles, the stellar flux was distributed on the image grid using the adaptive Gaussian smoothing scheme implemented in \texttt{pNbody}. Since a direct projection of the stellar particles would produce artificially noisy images in low-density regions, the flux of each particle was spread over neighbouring pixels with a Gaussian kernel whose characteristic scale was set by the local three-dimensional particle distribution, taken as the distance to the fifth-nearest stellar neighbour\footnote{We tested the sensitivity of the adaptive smoothing to the adopted neighbour number by repeating the mock image analysis for representative disrupted $M_\star=10^8$ and $10^9\,M_\odot$ systems using the sixth- and eighth-nearest stellar neighbours. In the surface brightness regime where extended debris is appreciably detected, the resulting $A_{\rm det}$ values show only small variations, while the qualitative morphology and detectability trends remain unchanged.}. This smoothing was used only to obtain a continuous representation of the projected stellar light from a discrete $N$-body realisation and is not intended to reproduce the instrumental point-spread function.

We quantified detectability through the projected area in arcsec\(^{2}\) of progenitor stellar light brighter than a nominal limiting surface brightness, \(\mu_{\rm lim}\), and denoted this quantity by \(A_{\rm det}\). In this work, \(\mu_{\rm lim}\) was not derived from an image noise model or from an \(n\sigma\) criterion within a given aperture, where \(\sigma\) denotes the standard deviation of the background fluctuations. Instead, it was imposed as a fixed surface brightness threshold, in units of \(\mathrm{mag\,arcsec^{-2}}\), applied to the idealised noiseless maps. We computed \(A_{\rm det}\) by summing the angular area of all map pixels with \(\mu \leq \mu_{\rm lim}\). Because all systems were placed at the same distance, this fixed distance sets the angular scale of the maps but does not affect the relative trends studied in this work.

At early times, when the progenitor is still intact, \(A_{\rm det}\) is dominated by the compact remnant; as stripping proceeds, increases in \(A_{\rm det}\) trace the emergence of a stream. Thus, \(A_{\rm det}\) should be understood as an image-based observability proxy, not as a dynamical decomposition of bound and stripped material.

\begin{figure*}[t]
    \centering
    \includegraphics[width=\textwidth]{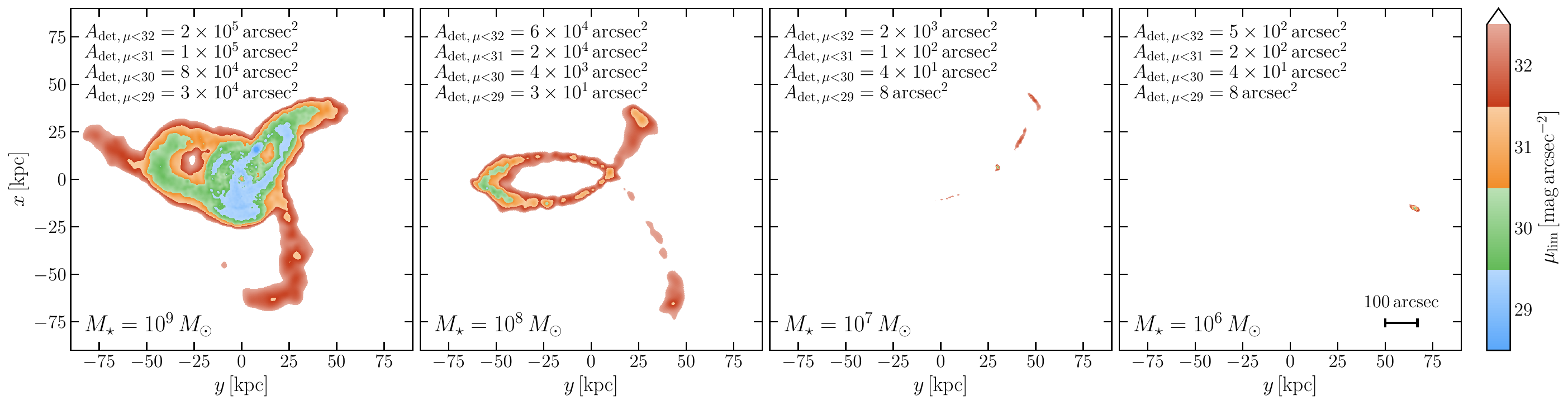}
    \caption{Examples of surface brightness maps for cuspy progenitors with stellar masses \(M_\star = 10^9, 10^8, 10^7,\) and \(10^6\,\mathrm{M_\odot}\), from left to right. The examples were selected at different orbits and evolutionary times to show visible tidal debris in each stellar mass case, and are intended to illustrate the connection between the integrated light maps and \(A_{\rm det}\), rather than to provide a controlled comparison at fixed orbit and time. The snapshots are shown at \(t = 3, 5.5, 4.5,\) and \(3\,\mathrm{Gyr}\), respectively, in the face-on projection. The maps were constructed in the ARRAKIHS VIS2 band assuming a distance of \(35\,\mathrm{Mpc}\), corresponding to a physical scale of \(100'' \simeq 17\,\mathrm{kpc}\). The colour scale encodes the surface brightness intervals adopted in this work, in units of \(\mathrm{mag\,arcsec^{-2}}\). The annotated values indicate the detectable area, \(A_{\rm det}\), associated with each limiting surface brightness threshold, as defined in Sect.~\ref{sec:sb_maps}. The coordinate origin corresponds to the centre of the host galaxy potential.}
    \label{fig:SB_example}
\end{figure*}

For each system, the projection was computed for two viewing angles relative to the initial orbital plane: a face-on projection (FO), perpendicular to the plane, and an edge-on projection (EO), parallel to it. This procedure allows us to assess the impact of projection effects on the apparent morphology and surface brightness of the tidal debris. Examples of the resulting surface brightness maps for different progenitor stellar masses are shown in Fig.~\ref{fig:SB_example}. These examples were selected to illustrate representative cases in which tidal debris is visible for each progenitor stellar mass. They are therefore not intended as a direct comparison at fixed orbit or fixed simulation time. Using the same orbit and snapshot for all stellar masses would in some cases show an intact progenitor and in others a more evolved stream, depending on the different stripping histories. Instead, the purpose of the figure is to illustrate how the integrated surface brightness maps are converted into the detectable area, \(A_{\rm det}\), and how the observable extent of the progenitor light changes with stellar mass and limiting surface brightness.

The mock images contain only the stellar particles associated with the dwarf galaxy progenitor. We did not include a luminous central galaxy, an extended stellar halo of the host, foreground or background contaminants, or instrumental systematics. Accordingly, the derived values of \(A_{\rm det}\) represent idealised upper limits set by the surface brightness threshold alone. In real observations, the recovery of diffuse tidal debris would also depend on the contrast with the central galaxy and the diffuse stellar halo of the host, foreground and background sources, and the adopted detection pipeline (see Infante-Sainz et al., in prep.). A full survey-specific image simulation is beyond the scope of this work.

We explored values in the range \(\mu_{\rm lim}=29\)--\(32\,\mathrm{mag\,arcsec^{-2}}\), chosen to bracket the sensitivity limits of current and forthcoming deep imaging surveys \citep[e.g.][]{lsst, Euclid2022}. The brighter end of this range is representative of wide-field surveys such as \textit{Euclid} and the LSST, while \(\mu_{\rm lim}\sim31\,\mathrm{mag\,arcsec^{-2}}\) is comparable to the design sensitivity of dedicated LSB missions such as ARRAKIHS.

\section{Results}\label{sec:results}
We examine how the detectability of tidal debris varies with orbital configuration, evolutionary stage, progenitor stellar mass, and inner dark matter density profile, while accounting for the effects of the adopted limiting surface brightness and projection. Following Sect.~\ref{sec:sb_maps}, detectability is measured from integrated surface brightness maps in the ARRAKIHS VIS2 band. For each nominal limiting surface brightness, \(\mu_{\rm lim}\), we define \(A_{\rm det}\) as the projected area of progenitor stellar light with \(\mu \leq \mu_{\rm lim}\). Unless otherwise stated, progenitors embedded in cuspy dark matter haloes were adopted as the fiducial case. We first analyse the dependence on orbital configuration and then consider the time evolution, stellar-mass dependence, and effect of the inner halo profile.
\subsection{Orbital dependence of tidal evolution and detectability}
\label{sec:results_orbit}

\subsubsection{Morphological classification across orbital configurations}
\label{subsec:results_morphology}

We first examine how orbital configuration affects the morphological state of the stellar component. At each simulation snapshot, we classified each system as intact, stream-like, or phase-mixed, motivated by the general distinction used in the FIRE and Auriga simulations of disrupted satellites \citep[e.g.][]{Riley_2025,Panithanpaisal_2021}. The classification was performed in two stages. We first separated intact progenitors from tidally disrupted candidates; the latter were classified as either stream-like or phase-mixed.

A physically motivated way to identify tidal disruption is to determine which stellar particles remain bound to the satellite and which have become unbound. We initially tested a binding-energy-based classification using the instantaneous Jacobi radius and the corresponding Jacobi-energy threshold \citep[see Chap.~8, Sect.~8.3 of][]{binney2008}. In this approach, the Jacobi radius defines an instantaneous tidal boundary around the satellite, set by its current position along the orbit. This boundary is associated with the saddle points of the effective potential, where the gravitational attraction of the satellite, the tidal field of the host, and the centrifugal term in the rotating frame balance. These saddle points act as escape channels from the satellite potential well. Particles outside the Jacobi radius are treated as unbound, while particles inside it are classified by comparing their Jacobi energy with the energy value at the saddle point. Particles with energies below this threshold are considered bound to the satellite.

However, for the eccentric orbits considered here, the instantaneous Jacobi radius is primarily set by the satellite's distance to the host. During close passages, the tidal boundary can contract sharply, increasing the fraction of particles classified as unbound even if the stellar component has not undergone a long-lived morphological transformation. The resulting bound fraction can be dominated by orbital phase rather than by the structural state of the progenitor, making it unsuitable as a stable threshold for separating intact from tidally disrupted systems.

We instead adopted a simpler aperture-based criterion, in which we first quantified the fraction of stellar particles that remain within a fixed spherical aperture of radius \(R_{\rm cut}\), centred on the tracked satellite centre, as \(f_{\rm in}(R_{\rm cut},t)=N_\star(r<R_{\rm cut},t)/N_\star(r<R_{\rm cut},0)\). We then used this quantity to define a morphological disruption criterion by applying a threshold to \(f_{\rm in}\). We tested a range of aperture radii, \(R_{\rm cut}\), and thresholds applied to \(f_{\rm in}\). The qualitative trends discussed below are unchanged by these choices. Smaller apertures, or larger thresholds in \(f_{\rm in}\), classify systems as tidally disrupted after a smaller amount of stellar material has moved away from the progenitor centre. We adopted \(R_{\rm cut}=15\,\mathrm{kpc}\) and a threshold of \(0.8\) applied to \(f_{\rm in}\) as our conservative fiducial definition. With this choice, models with \(f_{\rm in}<0.8\) are considered tidally disrupted candidates, while the remaining models are classified as intact. This aperture-based criterion provides a better correspondence with the visual distinction between intact and tidally disrupted systems in the simulation snapshots than the instantaneous Jacobi-based bound-fraction criterion.

After separating intact progenitors from tidally disrupted candidates, we classified the latter as stream-like or phase-mixed configurations using the local velocity structure of the stellar component. Stream-like systems are expected to remain dynamically colder, whereas phase-mixed systems have a more dispersed local velocity structure. Following the approach of \citet{Riley_2025}, we computed the median local three-dimensional velocity dispersion of all stellar particles associated with the simulated progenitor, \(\sigma_{\rm vel}^{50}\), and compared it with a stellar-mass-dependent threshold. We adopted their mass-dependent functional form but scaled the threshold by a factor of 1.2 to better match the morphologies identified visually in our simulations, such that \(\sigma_{\rm th}=1.2\,[3.51\log_{10}(M_\star/M_\odot)+1.08]\,\mathrm{km\,s^{-1}}\). The factor of 1.2 is therefore an empirical recalibration to our idealised simulations rather than a physically motivated constant. Tidally disrupted candidates with \(\sigma_{\rm vel}^{50}\geq\sigma_{\rm th}\) are classified as phase-mixed, whereas those below this threshold are classified as stream-like.

\begin{figure}[t]
    \centering
    \includegraphics[width=\linewidth]{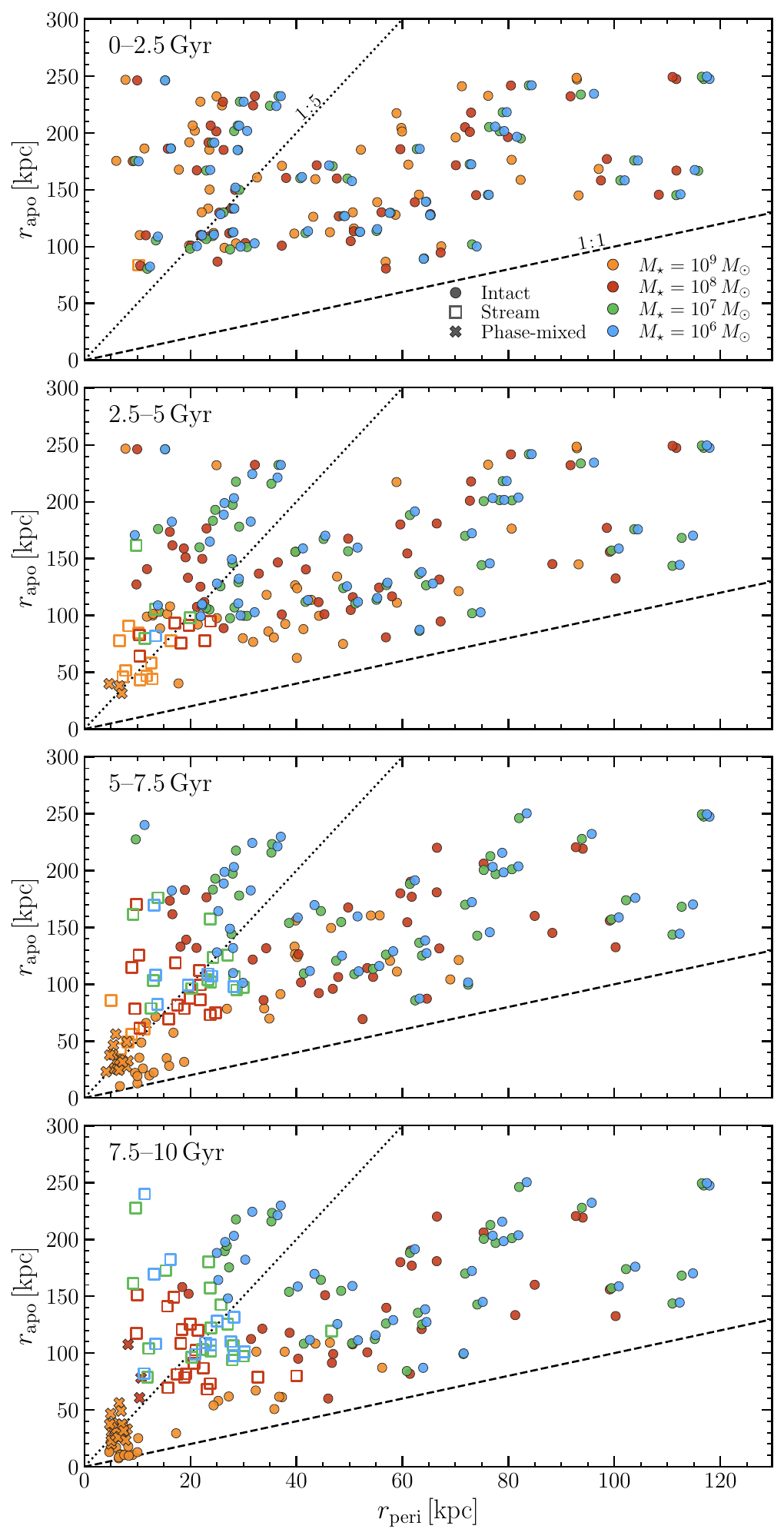}
    \caption{Distribution of orbital apocentres and pericentres for the fiducial cuspy models across four evolutionary stages. Each panel corresponds to a different \(2.5\,\mathrm{Gyr}\) time interval, and each point represents the apocentre--pericentre pair selected for a given orbit and progenitor stellar mass within that interval. Colours encode stellar mass, while symbols indicate the dominant classification state: intact (filled circles), stream-like (squares), or phase-mixed (stars), as defined in Sect.~\ref{sec:results_orbit}. The dashed and dotted black lines mark \(r_{\rm apo}/r_{\rm peri}=1\) and \(r_{\rm apo}/r_{\rm peri}=5\), respectively. The former corresponds to circular orbits, while the latter is shown as a reference for eccentric configurations similar to those occupied by several Milky Way satellites and streams \citep[e.g.][]{Shipp_2025}.}
    \label{fig:peri_apo_grid}
\end{figure}

Figure~\ref{fig:peri_apo_grid} shows the evolution of the orbital configurations of the progenitors in four \(2.5\,\mathrm{Gyr}\) time intervals after the start of the simulations. At early times, most models are classified as intact. During this first interval, points corresponding to different stellar masses remain close to one another because, for a given orbit, all progenitors were initialised with the same phase-space coordinates and mass-dependent orbital decay has not yet produced large separations. As the systems evolve, stream-like and phase-mixed classifications occur preferentially for progenitors that have evolved onto smaller radius orbits, with both small pericentres and small apocentres, corresponding to the lower-left region of the figure. In our orbital grid, the systems that reach this region correspond to the more eccentric initial configurations. This does not imply that tidal disruption is driven by eccentricity itself. Rather, disruption is mainly controlled by the amount of time spent in the inner regions of the host potential, where tidal forces are stronger. Our satellites were injected at large galactocentric distances to ensure that the simulations started before strong tidal processing by the host. As a result, satellites typically reach the inner regions of the host within the simulated time only if their orbits are sufficiently eccentric. This behaviour is clearest for the \(M_\star=10^8\) and \(10^9\,M_\odot\) progenitors, which develop extended tidal structures over the course of their evolution. The increasing separation between stellar masses in the pericentre--apocentre plane is partly driven by dynamical friction: more massive progenitors experience stronger orbital decay and therefore move to smaller orbital radii. However, their deeper potential wells and more tightly bound stellar components also make them more resistant to tidal stripping. As a result, orbital location alone does not determine the classification: massive progenitors can remain intact at smaller pericentres and apocentres than lower-mass systems, while less massive progenitors may be disrupted before reaching similarly small orbital radii.

Stream-like systems in our simulations are preferentially associated with small-pericentre orbits. Most stream-like classifications occur for \(r_{\rm peri}\lesssim30\)--\(40\,\mathrm{kpc}\), although a few cases extend to \(r_{\rm peri}\sim50\,\mathrm{kpc}\). These systems span a wider range of apocentric distances, especially at late times. The observed Milky Way dwarf galaxy streams also occupy small-pericentre orbits, with \(r_{\rm peri}<30\,\mathrm{kpc}\) for the sample analysed by \citet{2022Li}. Our simulations therefore produce stream-like systems in a similar region of the pericentre--apocentre plane. Similarly, some of our intact models survive after reaching relatively small pericentres, \(r_{\rm peri}\lesssim40\,\mathrm{kpc}\), overlapping with the orbital range occupied by Milky Way satellites \citep{Pace_2022}.

These results can be placed in the context of recent cosmological studies of disrupting satellites. \citet{Shipp_2023} find that detectable FIRE-2 streams extend to larger pericentres than the observed Milky Way dwarf galaxy streams, reaching \(r_{\rm peri}>110\,\mathrm{kpc}\), and survive as coherent streams only at larger apocentres. In Auriga, streams also occupy orbits comparable to those of observed Milky Way dwarf galaxy streams when the full sample of 28 haloes simulated at the lower resolution level 4 is considered, rather than only the higher resolution level 3 haloes adopted as the fiducial subset; several systems have \(r_{\rm peri}<25\,\mathrm{kpc}\) and \(r_{\rm apo}<100\,\mathrm{kpc}\), highlighting the importance of halo-to-halo variance \citep{Shipp_2025}. Observationally, streams around external galaxies are also often found at relatively small projected galactocentric distances. For example, the STReams in Residual Images of Nearby GalaxieS (STRRINGS) sample of long, narrow, curved streams shows that most of the selected streams lie within \(\sim30\,\mathrm{kpc}\) of their host galaxies \citep{Sola_2025}. Unlike the intact satellites in Auriga, which are generally found at large pericentres and apocentres, some of our models classified as intact can reach smaller pericentres, while remaining on comparable or extended apocentric orbits. In the \(M_\star=10^9\,\mathrm{M_\odot}\) cases with the smallest pericentres and apocentres, close to the host centre, a small amount of stellar material is stripped into a diffuse component while the main stellar body remains largely intact. These systems remain classified as intact because Fig.~\ref{fig:peri_apo_grid} shows the dominant morphology over each time interval and because they satisfy our dynamical criterion based on a fixed aperture, although diffuse stripped material may still be present. From an observational point of view, such material would also be difficult to recover because it is projected close to the central galaxy and would likely be affected by the host galaxy light.

\subsubsection{Detectable area across orbital configurations}

\begin{figure*}[t]
    \centering
    \includegraphics[width=\linewidth]{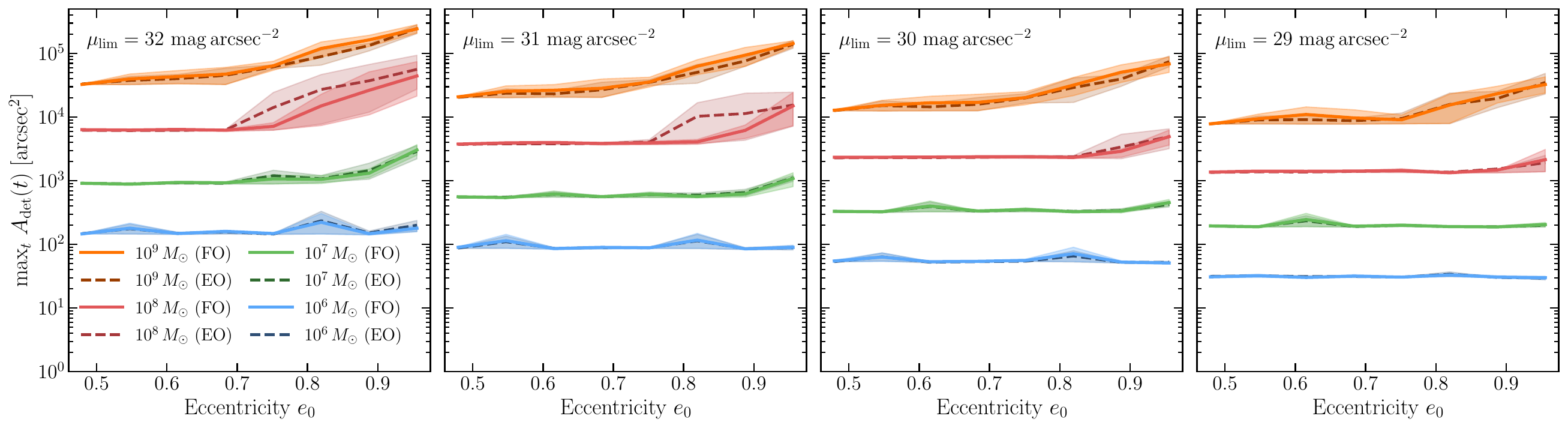}
    \caption{Maximum detectable progenitor-light area as a function of the initial orbital eccentricity proxy for the cuspy models, \(e_0\), defined in Sect.~\ref{sec:orbits}. Solid and dashed lines correspond to face-on and edge-on projections, respectively. For each stellar mass, the relation was computed in bins of \(e_0\): lines show the median value within each bin, while the shaded regions indicate the 16th--84th percentile range. Panels correspond to different surface brightness detection limits, \(\mu_{\rm lim} = 32, 31, 30,\) and \(29\,\mathrm{mag\,arcsec^{-2}}\).}
    \label{fig:Eccentricity}
\end{figure*}

To connect these morphological trends with observational detectability, we computed, for each system, the maximum detectable area reached over the full simulated evolution, \(\max_t A_{\rm det}(t)\). Here, \(A_{\rm det}(t)\) is the projected area of progenitor stellar light brighter than a given surface brightness detection limit, as defined in Sect.~\ref{sec:sb_maps}. Figure~\ref{fig:Eccentricity} shows the median \(\max_t A_{\rm det}(t)\) as a function of the initial eccentricity proxy, \(e_0\), for different values of \(\mu_{\rm lim}\) and progenitor stellar mass. 

The same orbital dependence identified in Fig.~\ref{fig:peri_apo_grid} is reflected in the detectability of the stellar material. The largest detectable areas are preferentially associated with the most eccentric initial configurations, because these are the systems that typically reach the inner regions of the host potential and spend sufficient time there for extended tidal structures to develop. In our simulations, the detectable area becomes significant mainly for \(e_0 \gtrsim 0.75\), particularly for the deepest limit, \(\mu_{\rm lim}=32\,\mathrm{mag\,arcsec^{-2}}\). Small local fluctuations in these relations mostly reflect the limited number of orbits per eccentricity bin and modest object-to-object variations in \(A_{\rm det}\), rather than separate physical regimes.

Figure~\ref{fig:Eccentricity} quantifies observability rather than disruption alone. Some low-mass progenitors are classified as tidally disrupted according to the criteria described above, but the stellar material removed from these systems remains too faint to exceed the adopted surface brightness limits. As a result, they contribute little or no detectable area despite undergoing stripping. The visibility of tidal debris is controlled jointly by the efficiency of tidal disruption, the stellar mass of the progenitor, and the limiting surface brightness of the observation.

Projection effects are comparatively weak. Although an edge-on view can slightly enhance the surface brightness by projecting more stars along the line of sight, the debris generally subtends a larger projected area in the face-on configuration. Since our detectability metric is based on area, these two effects largely compensate for each other, yielding similar trends for both projections. The main exception occurs for the \(M_\star = 10^8\,M_\odot\) models, for which the edge-on view begins to show detectable tidal structure at slightly lower eccentricities. Even in this case, the differences remain within the corresponding percentile ranges.

\subsection{Time evolution of the detectable area}\label{sec:result_time}

\begin{figure*}[t]
    \centering
    \includegraphics[width=\linewidth]{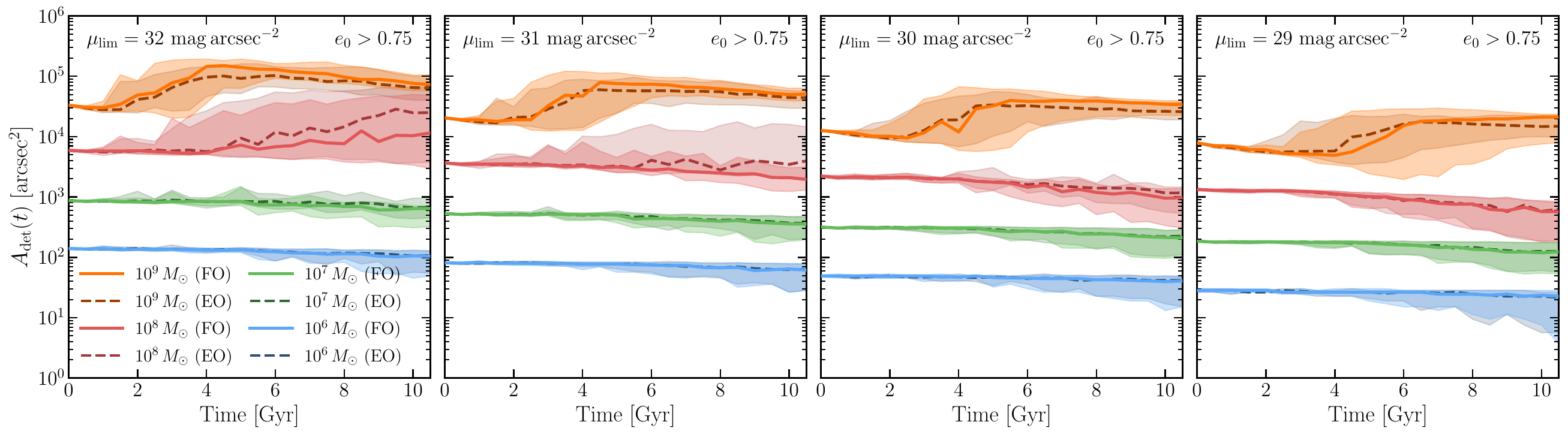}
    \caption{
    Time evolution of the detectable progenitor-light area for orbits with initial eccentricity \(e_0 > 0.75\) for the cuspy models. Solid and dashed lines correspond to face-on and edge-on projections, respectively. For each stellar mass, the curves show the median evolution across the selected orbits, while the shaded regions indicate the 16th--84th percentile range at each time step. Panels correspond to different surface brightness detection limits, \(\mu_{\rm lim} = 32, 31, 30,\) and \(29\,\mathrm{mag\,arcsec^{-2}}\).
    }
    \label{fig:Times}
\end{figure*}

Using the surface brightness maps generated at each simulation snapshot, we followed the time evolution of \(A_{\rm det}(t)\), i.e. the projected area of progenitor stellar light brighter than the adopted surface brightness detection limit. Figure~\ref{fig:Times} shows this quantity for the subset of orbits with high initial eccentricity proxy, \(e_0>0.75\), identified in the previous section as those producing the largest values of \(\max_t A_{\rm det}(t)\). 

For the lowest-mass progenitors, \(M_\star=10^6\) and \(10^7\,M_\odot\), \(A_{\rm det}(t)\) remains small and nearly constant over time. As stripping proceeds, part of the stellar material is removed from the central remnant, but the newly stripped debris is generally too diffuse to exceed the adopted surface brightness limits. The measured area can therefore remain nearly constant, or even decline, as the compact remnant loses stellar mass without a corresponding increase in detectable area. At late times, the percentile range broadens, reflecting differences in orbital phase and stripping history among the selected orbits. 

The \(M_\star=10^8\,M_\odot\) models show a clearer connection between stripping and detectability. In this case, the detectable area increases with time for the deeper limits, particularly for \(\mu_{\rm lim}=31\) and \(32\,\mathrm{mag\,arcsec^{-2}}\). At these limits, the debris produced after successive pericentric passages remains sufficiently bright to be recovered. At brighter surface brightness limits, much of the diffuse debris is fainter than the adopted limit, so the measured area remains dominated by the brighter remnant and the evolution resembles that of the lower-mass progenitors.

The most massive progenitors, \(M_\star=10^9\,M_\odot\), display detectable debris over a wider range of surface brightness limits. Their evolution is not monotonic: the detectable area initially increases as tidal stripping transforms the stellar component from an intact system into an extended stream-like structure, peaking at \(t\simeq4\)--\(6\,\mathrm{Gyr}\) after first infall, but decreases again at later times. This late-time decline is associated with the subsequent transition towards more phase-mixed morphologies. As the stellar material becomes more dispersed and the orbit decays towards the inner regions of the host potential, the projected debris distribution becomes more centrally concentrated around the host. Its spatial extent is reduced, lowering the detectable area despite the system being dynamically more disrupted. The offset in the time of the peak between different \(\mu_{\rm lim}\) values reflects the fact that deeper limits trace the extended diffuse debris as it forms, whereas brighter limits recover only the highest surface brightness parts of the stripped material and, therefore, peak later when the debris or the elongated remnant reaches a more detectable projected configuration.

\subsection{Stellar-mass dependence of the detectable area}\label{sec:results_mstar}

The dependence on progenitor stellar mass is already apparent in Figs.~\ref{fig:Eccentricity} and~\ref{fig:Times}, and it represents a key observational selection effect. At a fixed orbital configuration and evolutionary stage, increasing \(M_\star\) both increases the amount of stellar material available to be stripped and raises the surface brightness of the detectable progenitor light. The following trends refer mainly to the high-eccentricity orbits discussed above, \(e_0>0.75\), and to the evolutionary stages over which \(A_{\rm det}(t)\) is largest. In our simulations, extended debris is generally missed for \(M_\star=10^6\,M_\odot\) and barely detected for \(M_\star=10^7\,M_\odot\), even when tidal stripping occurs. The \(M_\star=10^8\,M_\odot\) models become detectable mainly at ARRAKIHS-like depths, \(\mu_{\rm lim}\simeq31\,\mathrm{mag\,arcsec^{-2}}\), and at the deepest limit considered here, \(\mu_{\rm lim}=32\,\mathrm{mag\,arcsec^{-2}}\). By contrast, the \(M_\star=10^9\,M_\odot\) progenitors produce the largest detectable areas and can still be recovered at brighter limits, \(\mu_{\rm lim}=29\) and \(30\,\mathrm{mag\,arcsec^{-2}}\). Thus, for a given orbital configuration, evolutionary stage, and surface brightness limit, the detectable area is dominated by the most massive progenitors within the explored grid.

\subsection{Inner-profile dependence of the detectable area}\label{sec:results_innerprofile}

As described in Sect.~\ref{sec:progenitors}, we also constructed cored versions of the fiducial progenitor models. The halo parameters were adjusted so that the resulting stellar velocity dispersions remained consistent with the observational constraints and comparable to those of the corresponding cuspy models. The stellar component was kept unchanged, allowing us to compare observationally matched cuspy and cored progenitors with similar stellar kinematics.

\begin{figure*}[t]
    \centering
    \includegraphics[width=\linewidth]{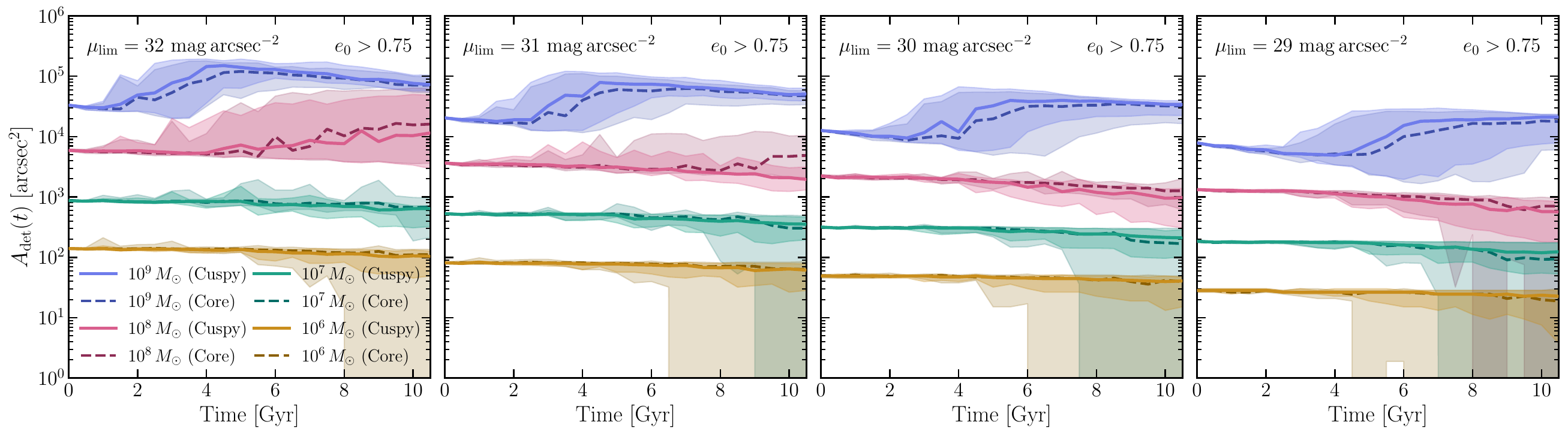}
    \caption{
    Time evolution of the detectable progenitor-light area for orbits with initial eccentricity $e_0 > 0.75$, shown for the face-on projection. Solid and dashed lines correspond to cuspy and cored models, respectively. For each stellar mass, the curves show the median evolution across the selected orbits, while the shaded regions indicate the 16th--84th percentile range at each time step. The four panels correspond to surface brightness detection limits of $\mu_{\rm lim} = 32, 31, 30,$ and $29\,\mathrm{mag\,arcsec^{-2}}$.
    }
    \label{fig:Time_core}
\end{figure*}

The cored progenitors were evolved using the same set of orbital initial conditions as the cuspy models. They follow the same qualitative dependence on orbital eccentricity as the cuspy progenitors, with the largest detectable areas preferentially produced by the high-eccentricity subset. We then compared their debris observability through the same metric used above: the projected area of stellar material brighter than a given nominal limiting surface brightness, \(\mu_{\rm lim}\). Figure~\ref{fig:Time_core} shows this comparison for the high-eccentricity subset, \(e_0>0.75\), in the face-on projection.

The main result is that cuspy and cored models show very similar detectable-area evolution. For all stellar masses, the median relations are close to each other and remain largely within the scatter associated with the different orbital configurations. This indicates that, within the parameter space explored here, orbital diversity introduces a larger variation in debris detectability than the difference between cuspy and cored inner profiles.

Some differences appear in the percentile ranges at late times. In particular, the lower envelope of the cored models can extend to smaller detectable areas than in the cuspy case for some stellar masses and detection limits. This behaviour is consistent with the lower central binding of cored progenitors, which makes their stellar component more susceptible to tidal stripping. In these more strongly stripped cases, part of the stellar material becomes too diffuse to remain detectable at the adopted limits, reducing the measured area. Conversely, cuspy systems tend to retain a compact bound remnant even after substantial mass loss, which can prevent the detectable area from vanishing entirely \citep{Errani_2019}. 

This analysis shows that the projected detectable area alone does not allow cuspy and cored progenitors to be clearly distinguished, because the differences associated with the cusp--core structure are subdominant to variations driven by orbital history, orbital phase, and surface brightness depth. A similar weak dependence on the assumed inner dark matter density profile has recently been found in idealised simulations of Milky Way dwarf spheroidals \citep{boyea2026}. 

\section{Discussion}
\label{sec:caveats}

The results presented in this work should be interpreted in the context of the idealised nature of the simulations and mock observations. The idealised setup is deliberately designed to isolate the role of progenitor stellar mass, orbital configuration, evolutionary stage, inner dark matter profile, projection, and surface brightness depth in shaping the detectability of tidal features. This controlled approach allows us to identify the main physical trends in \(A_{\rm det}\), while also defining the limitations that should be considered when interpreting the results and comparing them with future survey data.

First, the host galaxy was modelled as a static Milky Way-like potential, and orbital decay was included through an analytic Chandrasekhar dynamical-friction prescription. This approximation captures the global trend that more massive satellites experience stronger orbital decay, but it does not include the self-consistent response of a live host halo or the time evolution of the host disc, bulge, and dark matter halo. A natural extension of this work is therefore to repeat the analysis using live or explicitly time-dependent host potentials, together with a dynamical-friction treatment that evolves consistently with satellite mass loss.

Second, the simulated grid samples a finite region of parameter space. We explored four progenitor stellar masses, cuspy and cored dark matter haloes, and a broad set of bound orbital configurations, but we did not attempt to reproduce the full cosmological distribution of satellite infall times, orbital histories, stellar sizes, concentrations, or star-formation histories. In particular, the progenitor models were constructed as equilibrium systems without exploring variations in the stellar velocity anisotropy profile, which may affect the response of the stellar component to tidal stripping \citep{Read_2006, 2025MNRAS.544...36C}. Extending the grid to include additional progenitor structures and a wider range of orbital configurations will be required to build population-level predictions.

Third, the mock observations are idealised. The maps included only the progenitor stellar light and did not include host-galaxy light, an extended stellar halo, instrumental noise, background residuals, contaminating sources, point spread function effects, or a survey-specific detection pipeline. The detectable areas reported here should be regarded as upper limits set by the adopted surface brightness thresholds rather than as direct detection probabilities for a given survey. The brightest limit explored here, \(\mu_{\rm lim}=29\,\mathrm{mag\,arcsec^{-2}}\), already recovers little or no extended debris for the low-mass progenitors and mainly selects the highest surface brightness parts of the tidal structures. We consequently did not extend the analysis to brighter, shallower limits such as \(\mu_{\rm lim}=28\,\mathrm{mag\,arcsec^{-2}}\), which would further suppress the recovered debris and would not provide additional discriminatory power for the trends studied here. A direct comparison with facilities such as ARRAKIHS, \textit{Euclid}, or LSST will require survey-specific image simulations and recovery tests, including a minimum recoverable angular area and the effects of background subtraction and source contamination.

Finally, our analysis focuses primarily on the projected detectable area. This metric is useful for quantifying the amount of progenitor light above a given surface brightness limit, but it does not capture all information contained in the tidal debris. In particular, we find that \(A_{\rm det}\) alone does not clearly distinguish cuspy from cored progenitors, because the differences associated with the inner halo profile are smaller than the variations induced by orbital history, orbital phase, and surface brightness depth. Future work should combine the detectable area with additional diagnostics directly related to the disruption state, such as the survival of a bound progenitor remnant, the morphology of the tidal debris, multi-band colour information, and the stream width, to assess whether tidal features can provide stronger constraints on the inner dark matter profiles of dwarf galaxy progenitors.

\section{Conclusions}\label{sec:conclusion}
We used a controlled suite of 392 high-resolution, two-component $N$-body simulations to study the formation and observability of tidal debris produced by dwarf galaxies orbiting in a static Milky Way-like potential comprising a disc, a bulge, and a dark matter halo. The aim of this work is to quantify how the detectability of progenitor stellar light depends on progenitor stellar mass, orbital configuration, inner dark matter density profile, evolutionary stage, projection, and surface brightness depth. The simulations followed eight progenitor models, spanning \(M_\star=10^6\)--\(10^9\,M_\odot\) and including both cuspy and cored dark matter haloes, evolved for \(10\,\mathrm{Gyr}\) along 49 bound trajectories with initial eccentricity proxies in the range \(e_0\simeq0.45\)--\(0.99\). From idealised surface brightness maps in the ARRAKIHS VIS2 band, constructed for face-on and edge-on projections and for limiting surface brightnesses \(\mu_{\rm lim}=29\)--\(32\,\mathrm{mag\,arcsec^{-2}}\), we quantified detectability through the projected area of progenitor stellar light brighter than the adopted surface brightness limit, \(A_{\rm det}\). Our main conclusions are as follows:

\begin{enumerate}
    \item The formation of stream-like and phase-mixed stellar debris is strongly linked to orbital configuration. In the orbital range explored here, disrupted morphologies arise preferentially on eccentric, moderately radial orbits. When expressed in terms of observability, the largest detectable debris areas are associated with initial eccentricities \(e_0\gtrsim0.75\). This dependence on eccentricity reflects the fact that the progenitors were initially placed at large galactocentric distances, chosen to represent systems before strong tidal processing by the host. In this setup, high-\(e_0\) orbits are the configurations that bring the progenitors into the inner host potential for long enough to experience strong tidal stripping.

    \item Detectability is not equivalent to tidal disruption in our idealised surface brightness-limited maps. Several low-mass progenitors undergo stripping according to the dynamical and morphological criteria adopted in this work, but the stripped stellar material can remain too faint or too diffuse to be recovered at the surface brightness limits considered here. Within this framework, the absence of extended detectable debris does not necessarily indicate that a progenitor has remained dynamically intact, in agreement with similar conclusions from cosmological simulations \citep{2017MNRAS.468.4887W}.
    
    \item The detectable progenitor light area, \(A_{\rm det}(t)\), defined as the projected area of progenitor stellar light brighter than the adopted limiting surface brightness at time \(t\), does not always evolve monotonically. As objects evolve and lose more mass, stellar material is removed from the progenitor, but it also spreads spatially and can fade below the adopted surface brightness limits. For the \(M_\star=10^9\,M_\odot\) progenitors, \(A_{\rm det}(t)\) initially grows as stream-like structures form, typically reaching a maximum at \(t\simeq4\)--\(6\,\mathrm{Gyr}\) after first infall for the orbits with a high initial eccentricity proxy, \(e_0>0.75\), considered here, before decreasing at later times as the debris becomes more phase-mixed or more centrally concentrated around the host. In contrast, the \(M_\star=10^8\,M_\odot\) progenitors show a more gradual increase in \(A_{\rm det}(t)\) with time at the deepest surface brightness limits, as fewer of them reach the inner orbital regime where phase mixing and strong spatial diffusion reduce \(A_{\rm det}(t)\) at late times. Lower-mass progenitors, \(M_\star=10^6\)--\(10^7\,M_\odot\), show little growth in \(A_{\rm det}(t)\), since their stripped material generally falls below the adopted surface brightness limits.

    \item Progenitor stellar mass strongly affects the detectable area in our idealised surface brightness maps. This trend is evaluated at fixed surface brightness limits and within the orbital configurations, projections, and evolutionary stages explored here. Extended observable debris is generally missed for \(M_\star=10^6\,M_\odot\), for which the detectable area remains small, reaching at most \(A_{\rm det}\sim5\times10^2\,\mathrm{arcsec^2}\), and is mainly associated with the compact progenitor. Systems with \(M_\star=10^7\,M_\odot\) are only barely recovered, with detectable areas of order \(A_{\rm det}\sim2\times10^3\,\mathrm{arcsec^2}\) at the deepest limits. By contrast, \(M_\star=10^8\,M_\odot\) systems become detectable mainly at \(\mu_{\rm lim}\simeq31\)--\(32\,\mathrm{mag\,arcsec^{-2}}\), where the detectable area is typically of order \(A_{\rm det}\sim5\times10^4\,\mathrm{arcsec^2}\). The \(M_\star=10^9\,M_\odot\) progenitors dominate the detectable area, with values of order \(A_{\rm det}\sim10^5\,\mathrm{arcsec^2}\) at the deepest limits, and can still be recovered at brighter limits, \(\mu_{\rm lim}=29\)--\(30\,\mathrm{mag\,arcsec^{-2}}\). A direct prediction for a specific survey would require an observational selection function including noise, background subtraction, host-galaxy light, contaminants, and the detection pipeline.

    \item Projection effects are secondary compared to the dependence on orbit and stellar mass. Edge-on projections can increase the surface brightness by placing more stars along the line of sight, while face-on projections generally cover a larger apparent area. Since our metric is based on projected area, these effects largely compensate each other.

    \item Cuspy and cored progenitors show similar detectable-area evolution. We conclude that the projected detectable area alone does not allow the two inner dark matter profiles to be clearly distinguished, because the differences associated with the cusp--core structure are generally smaller than the scatter induced by orbital diversity. In this metric, the effect of the inner profile is largely degenerate with variations in orbital history, orbital phase, and surface brightness depth. Whether other diagnostics, such as detailed debris morphology, kinematics, multi-band colour information, or the spatial distribution of the stripped material, can provide stronger constraints on the inner dark matter profile remains to be explored.
\end{enumerate}
Future work will extend this controlled framework towards more realistic dynamical environments and a broader exploration of the information encoded in tidal debris. In particular, incorporating live or time-dependent host potentials and comparing our results with cosmological simulations will help establish which of the trends identified here persist in more complex environments and assembly histories. The progenitor models will also be extended to investigate the role of internal kinematics, particularly velocity anisotropy, in shaping the evolution and morphology of tidal debris. 
\begin{acknowledgements}

The research in this paper made use of the \textsc{SWIFT} open-source simulation code \citep{swift_code}, available at \url{http://www.swiftsim.com}, version 2025.04. Data analysis was carried out with the help of \texttt{pNbody} \citep{revaz_pnbody_2013}, available at \url{https://lastro.epfl.ch/projects/pNbody/}, as well as \textsc{NumPy} \citep{harris_numpy_2020} and \textsc{Matplotlib} \citep{Hunter_matplotlib_2007}. A. Pérez-Herrero acknowledges support from project PID2022-138896NB-C51 (MCIU/AEI/MINECO/FEDER, UE), funded by the Ministerio de Ciencia, Investigación y Universidades, and from a predoctoral fellowship awarded by the same ministry. STN and YR acknowledge the Swiss National Science Foundation (SNSF) under funding reference IZURZ2-224972 “Cosmic waltz of baryonic and ultralight dark matter: interaction and dynamical interplay”. We acknowledge EuroHPC Joint Undertaking for awarding us access to MareNostrum 5 at BSC, Spain, under the Extreme Scale Access Mode project EHPC-EXT-2025E01-023, and to the Red Española de Supercomputación (RES) to MareNostrum 5 at BSC, Spain, with projects AECT-2024-2-0008, AECT-2024-3-0002, AECT-2025-1-0007, and AECT-2025-2-0005. We acknowledge the Santander Supercomputing support group at the University of Cantabria for providing access to the Altamira Supercomputer at the Institute of Physics of Cantabria (IFCA-CSIC), a member of the Spanish Supercomputing Network, used to perform the simulations and analyses presented in this work. RB is supported by the SNSF through the Ambizione Grant PZ00P2-223532. APC is supported by a Yushan Fellowship awarded by the Taiwanese Ministry of Education, MOE-113-YSFMS-0002-001-P2 and acknowledges support from Taiwan's National Science and Technology Council (NSTC) under grant 113-2112-M-007-009. E.S. acknowledges support from the Leverhulme Trust Research Project Grant RPG-2021-205, "The Faint Universe Made Visible with Machine Learning". AF is supported by a Wallenberg Academy Fellowship. MAGF acknowledges financial support from the Spanish Ministry of Science and Innovation through the project PID2022-138896NB-C55. JP wishes to thank B. Sullivan, J. Sullivan and S. Sullivan for their inspiration. RIS acknowledges financial support from the Spanish Ministry of Science and Innovation (MCIN/AEI/10.13039/501100011033 y FEDER, Una manera de hacer Europa) with grant PID2022-138896NA-C54. PR acknowledges support from the Spanish Ministerio de Ciencia, Innovación y Universidades, through projects PID2022-138896NB; and the programme Unidad de Excelencia María de Maeztu, financed by MCIN/AEI/10.13039/501100011033, and by the MaX-CSIC Excellence Award MaX4-SOMMA-ICE. MA acknowledges project PID2024-162229NB-I00 financed by MICIU/AEI/10.13039/501100011033 and by FEDER, UE. The Instituto de Física de Cantabria (IFCA) has been accredited as Unidad de Excelencia María de Maeztu, a distinction awarded by the Spanish State Research Agency for the period 01/06/2026--31/05/2030. This work was supported by Fundação para a Ciência e a Tecnologia (FCT) through national funds under the research grant UID/04434/2025 (DOI: 10.54499/UID/04434/2025). P.P. acknowledges support from FCT through the grant 2024.17990.PEX and the Principal Investigator contract CIAAUP-092023-CTTI.

\end{acknowledgements}

\bibliographystyle{aa} 
\bibliography{bibliography}

@ARTICLE{bovy2015,
       author = {{Bovy}, Jo},
        title = {"{galpy: A python Library for Galactic Dynamics}"},
      journal = {\apjs},
         year = 2015,
        month = {02},
       volume = {216},
       number = {2},
          eid = {29},
        pages = {29},
          doi = {10.1088/0067-0049/216/2/29},
archivePrefix = {arXiv},
       eprint = {1412.3451},
 primaryClass = {astro-ph.GA},
       adsurl = {https://ui.adsabs.harvard.edu/abs/2015ApJS..216...29B}
}

@ARTICLE{revaz_2018,
       author = {{Revaz}, Yves and {Jablonka}, Pascale},
        title = "{Pushing back the limits: detailed properties of dwarf galaxies in a {\ensuremath{\Lambda}}CDM universe}",
      journal = {\aap},
         year = 2018,
        month = aug,
       volume = {616},
          eid = {A96},
        pages = {A96},
          doi = {10.1051/0004-6361/201832669},
archivePrefix = {arXiv},
       eprint = {1801.06222},
 primaryClass = {astro-ph.GA},
       adsurl = {https://ui.adsabs.harvard.edu/abs/2018A&A...616A..96R}
}

@MISC{swift_code,
  author = {{Schaller}, M. and others},
  title = "{SWIFT: SPH With Inter-dependent Fine-grained Tasking}",
  howpublished = {Astrophysics Source Code Library},
  year = 2018,
  month = may,
  eid = {ascl:1805.020},
  pages = {ascl:1805.020},
  archivePrefix = {ascl},
  eprint = {1805.020},
  adsurl = {https://ui.adsabs.harvard.edu/abs/2018ascl.soft05020S}
}

@MISC{revaz_pnbody_2013,
       author = {{Revaz}, Yves},
        title = "{pNbody: A python parallelized N-body reduction toolbox}",
 howpublished = {Astrophysics Source Code Library, record ascl:1302.004},
         year = {2013},
        month = {feb},
          eid = {ascl:1302.004},
       adsurl = {https://ui.adsabs.harvard.edu/abs/2013ascl.soft02004R}
}

@Article{         harris_numpy_2020,
 title         = {Array programming with {NumPy}},
 author        = {Charles R. Harris and K. Jarrod Millman and St{\'{e}}fan J.
                 van der Walt and Ralf Gommers and Pauli Virtanen and David
                 Cournapeau and Eric Wieser and Julian Taylor and Sebastian
                 Berg and Nathaniel J. Smith and Robert Kern and Matti Picus
                 and Stephan Hoyer and Marten H. van Kerkwijk and Matthew
                 Brett and Allan Haldane and Jaime Fern{\'{a}}ndez del
                 R{\'{i}}o and Mark Wiebe and Pearu Peterson and Pierre
                 G{\'{e}}rard-Marchant and Kevin Sheppard and Tyler Reddy and
                 Warren Weckesser and Hameer Abbasi and Christoph Gohlke and
                 Travis E. Oliphant},
 year          = {2020},
 month         = sep,
 journal       = {Nature},
 volume        = {585},
 number        = {7825},
 pages         = {357--362},
 doi           = {10.1038/s41586-020-2649-2},
 publisher     = {Springer Science and Business Media {LLC}},
 url           = {https://doi.org/10.1038/s41586-020-2649-2}
}

@Article{Hunter_matplotlib_2007,
  Author    = {Hunter, J. D.},
  Title     = {Matplotlib: A 2D graphics environment},
  Journal   = {Comput. Sci. Eng.},
  Volume    = {9},
  Number    = {3},
  Pages     = {90--95},
  publisher = {IEEE COMPUTER SOC},
  doi       = {10.1109/MCSE.2007.55},
  year      = 2007
}

@BOOK{binney2008,
       author = {{Binney}, James and {Tremaine}, Scott},
        title = "{Galactic Dynamics: Second Edition}",
         year = 2008,
       adsurl = {https://ui.adsabs.harvard.edu/abs/2008gady.book.....B}
}

@ARTICLE{vandenbosch2018b,
       author = {{van den Bosch}, Frank C. and {Ogiya}, Go and {Hahn}, Oliver and {Burkert}, Andreas},
        title = "{Disruption of dark matter substructure: fact or fiction?}",
      journal = {\mnras},
         year = 2018,
        month = mar,
       volume = {474},
       number = {3},
        pages = {3043-3066},
          doi = {10.1093/mnras/stx2956},
archivePrefix = {arXiv},
       eprint = {1711.05276},
 primaryClass = {astro-ph.GA},
       adsurl = {https://ui.adsabs.harvard.edu/abs/2018MNRAS.474.3043V}
}

@ARTICLE{vandenbosch2018a,
       author = {{van den Bosch}, Frank C. and {Ogiya}, Go},
        title = "{Dark matter substructure in numerical simulations: a tale of discreteness noise, runaway instabilities, and artificial disruption}",
      journal = {\mnras},
         year = 2018,
        month = apr,
       volume = {475},
       number = {3},
        pages = {4066-4087},
          doi = {10.1093/mnras/sty084},
archivePrefix = {arXiv},
       eprint = {1801.05427},
 primaryClass = {astro-ph.GA},
       adsurl = {https://ui.adsabs.harvard.edu/abs/2018MNRAS.475.4066V}
}

@ARTICLE{benson2022,
       author = {{Benson}, Andrew J. and {Du}, Xiaolong},
        title = "{Tidal tracks and artificial disruption of cold dark matter haloes}",
      journal = {\mnras},
         year = 2022,
        month = nov,
       volume = {517},
       number = {1},
        pages = {1398-1406},
          doi = {10.1093/mnras/stac2750},
archivePrefix = {arXiv},
       eprint = {2206.01842},
 primaryClass = {astro-ph.GA},
       adsurl = {https://ui.adsabs.harvard.edu/abs/2022MNRAS.517.1398B}
}

@article{Euclid2022,
   title={Euclidpreparation: XVI. Exploring the ultra-low surface brightness Universe withEuclid/VIS},
   volume={657},
   ISSN={1432-0746},
   url={http://dx.doi.org/10.1051/0004-6361/202141935},
   DOI={10.1051/0004-6361/202141935},
   journal={A\&A},
   publisher={EDP Sciences},
   author={Borlaff, A. S. and Gómez-Alvarez, P. and Altieri, B. and Marcum, P. M. and Vavrek, R. and Laureijs, R. and Kohley, R. and Buitrago, F. and Cuillandre, J.-C. and Duc, P.-A. and Gaspar Venancio, L. M. and Amara, A. and Andreon, S. and Auricchio, N. and Azzollini, R. and Baccigalupi, C. and Balaguera-Antolínez, A. and Baldi, M. and Bardelli, S. and Bender, R. and Biviano, A. and Bodendorf, C. and Bonino, D. and Bozzo, E. and Branchini, E. and Brescia, M. and Brinchmann, J. and Burigana, C. and Cabanac, R. and Camera, S. and Candini, G. P. and Capobianco, V. and Cappi, A. and Carbone, C. and Carretero, J. and Carvalho, C. S. and Casas, S. and Castander, F. J. and Castellano, M. and Castignani, G. and Cavuoti, S. and Cimatti, A. and Cledassou, R. and Colodro-Conde, C. and Congedo, G. and Conselice, C. J. and Conversi, L. and Copin, Y. and Corcione, L. and Coupon, J. and Courtois, H. M. and Cropper, M. and Da Silva, A. and Degaudenzi, H. and Di Ferdinando, D. and Douspis, M. and Dubath, F. and Duncan, C. A. J. and Dupac, X. and Dusini, S. and Ealet, A. and Fabricius, M. and Farina, M. and Farrens, S. and Ferreira, P. G. and Ferriol, S. and Finelli, F. and Flose-Reimberg, P. and Fosalba, P. and Frailis, M. and Franceschi, E. and Fumana, M. and Galeotta, S. and Ganga, K. and Garilli, B. and Gillis, B. and Giocoli, C. and Gozaliasl, G. and Graciá-Carpio, J. and Grazian, A. and Grupp, F. and Haugan, S. V. H. and Holmes, W. and Hormuth, F. and Jahnke, K. and Keihanen, E. and Kermiche, S. and Kiessling, A. and Kilbinger, M. and Kirkpatrick, C. C. and Kitching, T. and Knapen, J. H. and Kubik, B. and Kümmel, M. and Kunz, M. and Kurki-Suonio, H. and Liebing, P. and Ligori, S. and Lilje, P. B. and Lindholm, V. and Lloro, I. and Mainetti, G. and Maino, D. and Mansutti, O. and Marggraf, O. and Markovic, K. and Martinelli, M. and Martinet, N. and Martínez-Delgado, D. and Marulli, F. and Massey, R. and Maturi, M. and Maurogordato, S. and Medinaceli, E. and Mei, S. and Meneghetti, M. and Merlin, E. and Metcalf, R. B. and Meylan, G. and Moresco, M. and Morgante, G. and Moscardini, L. and Munari, E. and Nakajima, R. and Neissner, C. and Niemi, S. M. and Nightingale, J. W. and Nucita, A. and Padilla, C. and Paltani, S. and Pasian, F. and Patrizii, L. and Pedersen, K. and Percival, W. J. and Pettorino, V. and Pires, S. and Poncet, M. and Popa, L. and Potter, D. and Pozzetti, L. and Raison, F. and Rebolo, R. and Renzi, A. and Rhodes, J. and Riccio, G. and Romelli, E. and Roncarelli, M. and Rosset, C. and Rossetti, E. and Saglia, R. and Sánchez, A. G. and Sapone, D. and Sauvage, M. and Schneider, P. and Scottez, V. and Secroun, A. and Seidel, G. and Serrano, S. and Sirignano, C. and Sirri, G. and Skottfelt, J. and Stanco, L. and Starck, J. L. and Sureau, F. and Tallada-Crespí, P. and Taylor, A. N. and Tenti, M. and Tereno, I. and Teyssier, R. and Toledo-Moreo, R. and Torradeflot, F. and Tutusaus, I. and Valentijn, E. A. and Valenziano, L. and Valiviita, J. and Vassallo, T. and Viel, M. and Wang, Y. and Weller, J. and Whittaker, L. and Zacchei, A. and Zamorani, G. and Zucca, E.},
   year={2022},
   month=jan, pages={A92} }

@article{lsst,
    author = {Martin, G and Bazkiaei, A E and Spavone, M and Iodice, E and Mihos, J C and Montes, M and Benavides, J A and Brough, S and Carlin, J L and Collins, C A and Duc, P A and Gómez, F A and Galaz, G and Hernández-Toledo, H M and Jackson, R A and Kaviraj, S and Knapen, J H and Martínez-Lombilla, C and McGee, S and O’Ryan, D and Prole, D J and Rich, R M and Román, J and Shah, E A and Starkenburg, T K and Watkins, A E and Zaritsky, D and Pichon, C and Armus, L and Bianconi, M and Buitrago, F and Busá, I and Davis, F and Demarco, R and Desmons, A and García, P and Graham, A W and Holwerda, B and Hon, D S -H and Khalid, A and Klehammer, J and Klutse, D Y and Lazar, I and Nair, P and Noakes-Kettel, E A and Rutkowski, M and Saha, K and Sahu, N and Sola, E and Vázquez-Mata, J A and Vera-Casanova, A and Yoon, I},
    title = {Preparing for low surface brightness science with the Vera C. Rubin Observatory: Characterization of tidal features from mock images},
    journal = {MNRAS},
    volume = {513},
    number = {1},
    pages = {1459-1487},
    year = {2022},
    month = {04},
    issn = {0035-8711},
    doi = {10.1093/mnras/stac1003},
    url = {https://doi.org/10.1093/mnras/stac1003},
    eprint = {https://academic.oup.com/mnras/article-pdf/513/1/1459/43515818/stac1003.pdf},
}

@article{Agertz_2021,
   title={VINTERGATAN – I. The origins of chemically, kinematically, and structurally distinct discs in a simulated Milky Way-mass galaxy},
   volume={503},
   ISSN={1365-2966},
   url={http://dx.doi.org/10.1093/mnras/stab322},
   DOI={10.1093/mnras/stab322},
   number={4},
   journal={MNRAS},
   publisher={Oxford University Press (OUP)},
   author={Agertz, Oscar and Renaud, Florent and Feltzing, Sofia and Read, Justin I and Ryde, Nils and Andersson, Eric P and Rey, Martin P and Bensby, Thomas and Feuillet, Diane K},
   year={2021},
   month=apr, pages={5826–5845} }

@article{Errani_2019,
   title={Can tides disrupt cold dark matter subhaloes?},
   volume={491},
   ISSN={1365-2966},
   url={http://dx.doi.org/10.1093/mnras/stz3349},
   DOI={10.1093/mnras/stz3349},
   number={4},
   journal={MNRAS},
   publisher={Oxford University Press (OUP)},
   author={Errani, Raphaël and Peñarrubia, Jorge},
   year={2019},
   month=dec, pages={4591–4601} }

@article{Errani_2022,
   title={Dark matter halo cores and the tidal survival of Milky Way satellites},
   volume={519},
   ISSN={1365-2966},
   url={http://dx.doi.org/10.1093/mnras/stac3499},
   DOI={10.1093/mnras/stac3499},
   number={1},
   journal={MNRAS},
   publisher={Oxford University Press (OUP)},
   author={Errani, Raphaël and Navarro, Julio F and Peñarrubia, Jorge and Famaey, Benoit and Ibata, Rodrigo},
   year={2022},
   month=nov, pages={384–396} }

@article{Navarro_1997,
   title={A Universal Density Profile from Hierarchical Clustering},
   volume={490},
   ISSN={1538-4357},
   url={http://dx.doi.org/10.1086/304888},
   DOI={10.1086/304888},
   number={2},
   journal={ApJ},
   publisher={American Astronomical Society},
   author={Navarro, Julio F. and Frenk, Carlos S. and White, Simon D. M.},
   year={1997},
   month=dec, pages={493–508} }

@article{Sola_2025,
   title={STRRINGS: STReams in Residual Images of Nearby GalaxieS},
   volume={544},
   ISSN={1365-2966},
   url={http://dx.doi.org/10.1093/mnras/staf1647},
   DOI={10.1093/mnras/staf1647},
   number={1},
   journal={MNRAS},
   publisher={Oxford University Press (OUP)},
   author={Sola, Elisabeth and Chemaly, David and Belokurov, Vasily and Müller, Oliver and Ardern-Arentsen, Anke and Davies, Elliot Y and Laguna-Miralles, Júlia and Myeong, GyuChul and Panagiotakis, Konstantinos and Zhang, HanYuan and Erkal, Denis and Koposov, Sergey E and Lang, Dustin and Nibauer, Jacob},
   year={2025},
   month=sep, pages={735–763} }

@article{Erkal_2016,
   title={The number and size of subhalo-induced gaps in stellar streams},
   volume={463},
   ISSN={1365-2966},
   url={http://dx.doi.org/10.1093/mnras/stw1957},
   DOI={10.1093/mnras/stw1957},
   number={1},
   journal={MNRAS},
   publisher={Oxford University Press (OUP)},
   author={Erkal, Denis and Belokurov, Vasily and Bovy, Jo and Sanders, Jason L.},
   year={2016},
   month=aug, pages={102–119} }

@misc{stoiber2025,
      title={Living the stream: Properties and progenitors of tidal shells and streams around galaxies from Magneticum}, 
      author={Johannes Stoiber and Lucas M. Valenzuela and Rhea-Silvia Remus and Lucas C. Kimmig and Jan-Niklas Pippert and Elisabeth Sola and Klaus Dolag},
      year={2025},
      eprint={2509.25307},
      archivePrefix={arXiv},
      primaryClass={astro-ph.GA},
      url={https://arxiv.org/abs/2509.25307}, 
}

@article{Shipp_2025,
   title={Auriga Streams II: orbital properties of tidally disrupting satellites of Milky Way-mass galaxies},
   volume={542},
   ISSN={1365-2966},
   url={http://dx.doi.org/10.1093/mnras/staf1283},
   DOI={10.1093/mnras/staf1283},
   number={2},
   journal={MNRAS},
   publisher={Oxford University Press (OUP)},
   author={Shipp, Nora and Riley, Alexander H and Simpson, Christine M and Bieri, Rebekka and Necib, Lina and Arora, Arpit and Fragkoudi, Francesca and Gómez, Facundo A and Grand, Robert J J and Marinacci, Federico},
   year={2025},
   month=aug, pages={1109–1124} }

@article{Riley_2025,
   title={Auriga Streams – I: disrupting satellites surrounding Milky Way-mass haloes at multiple resolutions},
   volume={542},
   ISSN={1365-2966},
   url={http://dx.doi.org/10.1093/mnras/staf1350},
   DOI={10.1093/mnras/staf1350},
   number={3},
   journal={MNRAS},
   publisher={Oxford University Press (OUP)},
   author={Riley, Alexander H and Shipp, Nora and Simpson, Christine M and Bieri, Rebekka and Fattahi, Azadeh and Brown, Shaun T and Oman, Kyle A and Fragkoudi, Francesca and Gómez, Facundo A and Grand, Robert J J and Marinacci, Federico},
   year={2025},
   month=aug, pages={2443–2463} }

@article{Shipp_2023,
   title={Streams on FIRE: Populations of Detectable Stellar Streams in the Milky Way and FIRE},
   volume={949},
   ISSN={1538-4357},
   url={http://dx.doi.org/10.3847/1538-4357/acc582},
   DOI={10.3847/1538-4357/acc582},
   number={2},
   journal={ApJ},
   publisher={American Astronomical Society},
   author={Shipp, Nora and Panithanpaisal, Nondh and Necib, Lina and Sanderson, Robyn and Erkal, Denis and Li, Ting S. and Santistevan, Isaiah B. and Wetzel, Andrew and Cullinane, Lara R. and Ji, Alexander P. and Koposov, Sergey E. and Kuehn, Kyler and Lewis, Geraint F. and Pace, Andrew B. and Zucker, Daniel B. and Bland-Hawthorn, Joss and Cunningham, Emily C. and Kim, Stacy Y. and Lilleengen, Sophia and Moreno, Jorge and Sharma, Sanjib},
   year={2023},
   month=may, pages={44} }

@article{Grand_2017,
   title={The Auriga Project: the properties and formation mechanisms of disc galaxies across cosmic time},
   ISSN={1365-2966},
   url={http://dx.doi.org/10.1093/mnras/stx071},
   DOI={10.1093/mnras/stx071},
   journal={MNRAS},
   publisher={Oxford University Press (OUP)},
   author={Grand, Robert J. J. and Gómez, Facundo A. and Marinacci, Federico and Pakmor, Rüdiger and Springel, Volker and Campbell, David J. R. and Frenk, Carlos S. and Jenkins, Adrian and White, Simon D. M.},
   year={2017},
   month=jan, pages={stx071} }

@article{Panithanpaisal_2021,
   title={The Galaxy Progenitors of Stellar Streams around Milky Way–mass Galaxies in the FIRE Cosmological Simulations},
   volume={920},
   ISSN={1538-4357},
   url={http://dx.doi.org/10.3847/1538-4357/ac1109},
   DOI={10.3847/1538-4357/ac1109},
   number={1},
   journal={ApJ},
   publisher={American Astronomical Society},
   author={Panithanpaisal, Nondh and Sanderson, Robyn E. and Wetzel, Andrew and Cunningham, Emily C. and Bailin, Jeremy and Faucher-Giguère, Claude-André},
   year={2021},
   month=oct, pages={10} }

@ARTICLE{StreamGen,
       author = {{Dropulic}, Adriana and {Shipp}, Nora and {Kim}, Stacy and {Mezghanni}, Zeineb and {Necib}, Lina and {Lisanti}, Mariangela},
        title = "{StreamGen: Connecting Populations of Streams and Shells to Their Host Galaxies}",
      journal = {\apj},
         year = 2025,
        month = sep,
       volume = {990},
       number = {2},
          eid = {162},
        pages = {162},
          doi = {10.3847/1538-4357/adf1a7},
archivePrefix = {arXiv},
       eprint = {2409.13810},
 primaryClass = {astro-ph.GA},
       adsurl = {https://ui.adsabs.harvard.edu/abs/2025ApJ...990..162D}
}

@article{Mir_Carretero_2024,
   title={Extragalactic stellar tidal streams in the Dark Energy Survey},
   volume={691},
   ISSN={1432-0746},
   url={http://dx.doi.org/10.1051/0004-6361/202451685},
   DOI={10.1051/0004-6361/202451685},
   journal={A\&A},
   publisher={EDP Sciences},
   author={Miró-Carretero, Juan and Martínez-Delgado, David and Gómez-Flechoso, María A. and Cooper, Andrew and Akhlaghi, Mohammad and Donatiello, Giuseppe and Kuijken, Konrad and Lang, Dustin and Makarov, Dmitry and Laine, Seppo and Roca-Fàbrega, Santi},
   year={2024},
   month=nov, pages={A196} }

@misc{arias_2025,
      title={Energy-Space Analysis of Tidal Stripping in Stellar-Dark Matter Systems}, 
      author={Bradley Arias and Nicole E. Drakos and James E. Taylor},
      year={2025},
      eprint={2508.16759},
      archivePrefix={arXiv},
      primaryClass={astro-ph.GA},
      url={https://arxiv.org/abs/2508.16759}, 
}

@article{Aguirre_2025,
   title={New insights on low-mass dark matter subhalo tidal tracks via numerical simulations},
   volume={545},
   ISSN={1365-2966},
   url={http://dx.doi.org/10.1093/mnras/staf2058},
   DOI={10.1093/mnras/staf2058},
   number={2},
   journal={MNRAS},
   publisher={Oxford University Press (OUP)},
   author={Aguirre-Santaella, Alejandra and Sánchez-Conde, Miguel A and Ogiya, Go},
   year={2025},
   month=nov }

@ARTICLE{Plummer1911,
       author = {{Plummer}, H.~C.},
        title = "{On the problem of distribution in globular star clusters}",
      journal = {\mnras},
         year = 1911,
        month = mar,
       volume = {71},
        pages = {460-470},
          doi = {10.1093/mnras/71.5.460},
       adsurl = {https://ui.adsabs.harvard.edu/abs/1911MNRAS..71..460P}
}

@article{Zhao_1996,
   title={Analytical models for galactic nuclei},
   volume={278},
   ISSN={1365-2966},
   url={http://dx.doi.org/10.1093/mnras/278.2.488},
   DOI={10.1093/mnras/278.2.488},
   number={2},
   journal={MNRAS},
   publisher={Oxford University Press (OUP)},
   author={Zhao, H.},
   year={1996},
   month=jan, pages={488–496} }

@article{Diemer_2013,
   title={THE PSEUDO-EVOLUTION OF HALO MASS},
   volume={766},
   ISSN={1538-4357},
   url={http://dx.doi.org/10.1088/0004-637X/766/1/25},
   DOI={10.1088/0004-637x/766/1/25},
   number={1},
   journal={ApJ},
   publisher={American Astronomical Society},
   author={Diemer, Benedikt and More, Surhud and Kravtsov, Andrey V.},
   year={2013},
   month=mar, pages={25} }

@article{McConnachie_2020,
   title={Solo dwarfs – III. Exploring the orbital origins of isolated Local Group galaxies with Gaia Data Release 2},
   volume={501},
   ISSN={1365-2966},
   url={http://dx.doi.org/10.1093/mnras/staa3740},
   DOI={10.1093/mnras/staa3740},
   number={2},
   journal={MNRAS},
   publisher={Oxford University Press (OUP)},
   author={McConnachie, Alan W and Higgs, Clare R and Thomas, Guillaume F and Venn, Kim A and Côté, Patrick and Battaglia, Giuseppina and Lewis, Geraint F},
   year={2020},
   month=dec, pages={2363–2377} }

@misc{elves,
      title={ELVES-Field: Isolated Dwarf Galaxy Quenched Fractions Rise Below $M_* \approx 10^7$ $M_\odot$}, 
      author={Scott Carlsten and Jiaxuan Li and Jenny Greene and Alex Drlica-Wagner and Shany Danieli},
      year={2026},
      eprint={2602.16778},
      archivePrefix={arXiv},
      primaryClass={astro-ph.GA},
      url={https://arxiv.org/abs/2602.16778}, 
}

@ARTICLE{2022Li,
       author = {{Li}, Ting S. and {Ji}, Alexander P. and {Pace}, Andrew B. and {Erkal}, Denis and {Koposov}, Sergey E. and {Shipp}, Nora and {Da Costa}, Gary S. and {Cullinane}, Lara R. and {Kuehn}, Kyler and {Lewis}, Geraint F. and {Mackey}, Dougal and {Simpson}, Jeffrey D. and {Zucker}, Daniel B. and {Ferguson}, Peter S. and {Martell}, Sarah L. and {Bland-Hawthorn}, Joss and {Balbinot}, Eduardo and {Tavangar}, Kiyan and {Drlica-Wagner}, Alex and {De Silva}, Gayandhi M. and {Simon}, Joshua D.},
        title = "{S $^{5}$: The Orbital and Chemical Properties of One Dozen Stellar Streams}",
      journal = {\apj},
         year = 2022,
        month = mar,
       volume = {928},
       number = {1},
          eid = {30},
        pages = {30},
          doi = {10.3847/1538-4357/ac46d3},
archivePrefix = {arXiv},
       eprint = {2110.06950},
 primaryClass = {astro-ph.GA},
       adsurl = {https://ui.adsabs.harvard.edu/abs/2022ApJ...928...30L}
}

@article{Dutton:2014xda,
    author = "Dutton, Aaron A. and Macci{\`o}, Andrea V.",
    title = "{Cold dark matter haloes in the Planck era: evolution of structural parameters for Einasto and NFW profiles}",
    eprint = "1402.7073",
    archivePrefix = "arXiv",
    primaryClass = "astro-ph.CO",
    doi = "10.1093/mnras/stu742",
    journal = {MNRAS},
    volume = "441",
    number = "4",
    pages = "3359--3374",
    year = "2014"
}

@ARTICLE{Pace_2022,
       author = {{Pace}, Andrew B. and {Erkal}, Denis and {Li}, Ting S.},
        title = "{Proper Motions, Orbits, and Tidal Influences of Milky Way Dwarf Spheroidal Galaxies}",
      journal = {\apj},
         year = 2022,
        month = dec,
       volume = {940},
       number = {2},
          eid = {136},
        pages = {136},
          doi = {10.3847/1538-4357/ac997b},
archivePrefix = {arXiv},
       eprint = {2205.05699},
 primaryClass = {astro-ph.GA},
       adsurl = {https://ui.adsabs.harvard.edu/abs/2022ApJ...940..136P}
}

@ARTICLE{byrne2022,
       author = {{Byrne}, C.~M. and {Stanway}, E.~R. and {Eldridge}, J.~J. and {McSwiney}, L. and {Townsend}, O.~T.},
        title = "{The dependence of theoretical synthetic spectra on {\ensuremath{\alpha}}-enhancement in young, binary stellar populations}",
      journal = {\mnras},
         year = 2022,
        month = jun,
       volume = {512},
       number = {4},
        pages = {5329-5338},
          doi = {10.1093/mnras/stac807},
archivePrefix = {arXiv},
       eprint = {2203.13275},
 primaryClass = {astro-ph.SR},
       adsurl = {https://ui.adsabs.harvard.edu/abs/2022MNRAS.512.5329B}
}

@article{Trelles_2022,
   title={Concurrent infall of satellites: Collective effects changing the overall picture},
   volume={668},
   ISSN={1432-0746},
   url={http://dx.doi.org/10.1051/0004-6361/202243671},
   DOI={10.1051/0004-6361/202243671},
   journal = {A\&A},
   publisher={EDP Sciences},
   author={Trelles, A. and Valenzuela, O. and Roca-Fábrega, S. and Velázquez, H.},
   year={2022},
   month=Nov, pages={A20} }

@ARTICLE{casertano,
       author = {{Casertano}, S. and {Hut}, P.},
        title = "{Core radius and density measurements in N-body experiments Connections with theoretical and observational definitions}",
      journal = {\apj},
         year = 1985,
        month = nov,
       volume = {298},
        pages = {80-94},
          doi = {10.1086/163589},
       adsurl = {https://ui.adsabs.harvard.edu/abs/1985ApJ...298...80C}
}

@article{ refId0,
	author = {{Euclid Collaboration} and {Cropper, M. S.} and {Al-Bahlawan, A.} and {Amiaux, J.} and {Awan, S.} and {Azzollini, R.} and {Benson, K.} and {Berthe, M.} and {Boucher, J.} and {Bozzo, E.} and {Brockley-Blatt, C.} and {Candini, G. P.} and {Cara, C.} and {Chaudery, R. A.} and {Cole, R. E.} and {Danto, P.} and {Denniston, J.} and {Di Giorgio, A. M.} and {Dryer, B.} and {Dubois, J.-P.} and {Endicott, J.} and {Farina, M.} and {Galli, E.} and {Genolet, L.} and {Gow, J. P. D.} and {Guttridge, P.} and {Hailey, M.} and {Hall, D.} and {Harper, C.} and {Hoekstra, H.} and {Holland, A. D.} and {Horeau, B.} and {Hu, D.} and {James, R. E.} and {Khalil, A.} and {King, R.} and {Kitching, T.} and {Kohley, R.} and {Larcheveque, C.} and {Lawrenson, A.} and {Liebing, P.} and {Liu, S. J.} and {Martignac, J.} and {Massey, R.} and {McCracken, H. J.} and {Miller, L.} and {Murray, N.} and {Nakajima, R.} and {Niemi, S.-M.} and {Nightingale, J. W.} and {Paltani, S.} and {Pendem, A.} and {Philippon, A.} and {Plana, C.} and {Pool, P.} and {Pottinger, S.} and {Racca, G. D.} and {Rhodes, J.} and {Rousseau, A.} and {Ruane, K.} and {Salatti, M.} and {Salvignol, J.-C.} and {Sciortino, A.} and {Short, A.} and {Skottfelt, J.} and {Smit, S. J. A.} and {Swindells, I.} and {Szafraniec, M.} and {Thomas, P. D.} and {Thomas, W.} and {Tommasi, E.} and {Tosti, S.} and {Visticot, F.} and {Walton, D. M.} and {Willis, G.} and {Winter, B.} and {Aghanim, N.} and {Altieri, B.} and {Amara, A.} and {Andreon, S.} and {Auricchio, N.} and {Aussel, H.} and {Baccigalupi, C.} and {Baldi, M.} and {Balestra, A.} and {Bardelli, S.} and {Basset, A.} and {Bender, R.} and {Bernardeau, F.} and {Bodendorf, C.} and {Boenke, T.} and {Bonino, D.} and {Branchini, E.} and {Brescia, M.} and {Brinchmann, J.} and {Camera, S.} and {Capobianco, V.} and {Carbone, C.} and {Cardone, V. F.} and {Carretero, J.} and {Casas, R.} and {Casas, S.} and {Castander, F. J.} and {Castellano, M.} and {Castignani, G.} and {Cavuoti, S.} and {Cimatti, A.} and {Colodro-Conde, C.} and {Congedo, G.} and {Conselice, C. J.} and {Conversi, L.} and {Copin, Y.} and {Courbin, F.} and {Courtois, H. M.} and {Crocce, M.} and {Cuby, J.-G.} and {Cuillandre, J.-C.} and {Da Silva, A.} and {Degaudenzi, H.} and {De Lucia, G.} and {Dinis, J.} and {Dolding, C.} and {Douspis, M.} and {Duncan, C. A. J.} and {Dupac, X.} and {Dusini, S.} and {Ealet, A.} and {Fabricius, M.} and {Farrens, S.} and {Ferriol, S.} and {Fosalba, P.} and {Fotopoulou, S.} and {Frailis, M.} and {Franceschi, E.} and {Franzetti, P.} and {Frugier, P.-A.} and {Fumana, M.} and {Galeotta, S.} and {Garilli, B.} and {George, K.} and {Gillard, W.} and {Gillis, B.} and {Giocoli, C.} and {G\'omez-Alvarez, P.} and {Granett, B. R.} and {Grazian, A.} and {Grupp, F.} and {Guzzo, L.} and {Haugan, S. V. H.} and {Herent, O.} and {Hoar, J.} and {Holliman, M. S.} and {Holmes, W.} and {Hook, I.} and {Hormuth, F.} and {Hornstrup, A.} and {Hudelot, P.} and {Ili\'{}c, S.} and {Jahnke, K.} and {Jhabvala, M.} and {Joachimi, B.} and {Keih\"anen, E.} and {Kermiche, S.} and {Kiessling, A.} and {Kilbinger, M.} and {Kubik, B.} and {Kuijken, K.} and {K\"ummel, M.} and {Kunz, M.} and {Kurki-Suonio, H.} and {Lahav, O.} and {Laureijs, R.} and {Ligori, S.} and {Lilje, P. B.} and {Lindholm, V.} and {Lloro, I.} and {Lorenzo Alvarez, J.} and {Mainetti, G.} and {Maino, D.} and {Maiorano, E.} and {Mansutti, O.} and {Marcin, S.} and {Marggraf, O.} and {Markovic, K.} and {Martinelli, M.} and {Martinet, N.} and {Marulli, F.} and {Masters, D. C.} and {Maurogordato, S.} and {Medinaceli, E.} and {Mei, S.} and {Melchior, M.} and {Mellier, Y.} and {Meneghetti, M.} and {Merlin, E.} and {Meylan, G.} and {Mohr, J. J.} and {Moresco, M.} and {Moscardini, L.} and {Neissner, C.} and {Nichol, R. C.} and {Nutma, T.} and {Padilla, C.} and {Paech, K.} and {Pasian, F.} and {Peacock, J. A.} and {Pedersen, K.} and {Percival, W. J.} and {Pettorino, V.} and {Pires, S.} and {Polenta, G.} and {Poncet, M.} and {Popa, L. A.} and {Pozzetti, L.} and {Raison, F.} and {Rebolo, R.} and {Refregier, A.} and {Renzi, A.} and {Riccio, G.} and {Rix, Hans-Walter} and {Romelli, E.} and {Roncarelli, M.} and {Rosset, C.} and {Rossetti, E.} and {Rottgering, H. J. A.} and {Rusholme, B.} and {Saglia, R.} and {Sakr, Z.} and {S\'anchez, A. G.} and {Sapone, D.} and {Sauvage, M.} and {Scaramella, R.} and {Schewtschenko, J. A.} and {Schirmer, M.} and {Schneider, P.} and {Schrabback, T.} and {Secroun, A.} and {Sefusatti, E.} and {Seidel, G.} and {Seiffert, M.} and {Serrano, S.} and {Sirignano, C.} and {Sirri, G.} and {Stanco, L.} and {Starck, J.-L.} and {Steinwagner, J.} and {Tallada-Cresp\'{\i}, P.} and {Tavagnacco, D.} and {Taylor, A. N.} and {Teplitz, H. I.} and {Tereno, I.} and {Toledo-Moreo, R.} and {Torradeflot, F.} and {Tutusaus, I.} and {Valentijn, E. A.} and {Valenziano, L.} and {Vassallo, T.} and {Verdoes Kleijn, G.} and {Veropalumbo, A.} and {Wachter, S.} and {Wang, Y.} and {Weller, J.} and {Zamorani, G.} and {Zoubian, J.} and {Zucca, E.} and {Biviano, A.} and {Bolzonella, M.} and {Boucaud, A.} and {Burigana, C.} and {Calabrese, M.} and {Casenove, P.} and {Di Ferdinando, D.} and {Escartin Vigo, J. A.} and {Fabbian, G.} and {Farinelli, R.} and {Finelli, F.} and {Gracia-Carpio, J.} and {Israel, H.} and {Mauri, N.} and {Nguyen-Kim, H. N.} and {Pezzotta, A.} and {P\"ontinen, M.} and {Porciani, C.} and {Scottez, V.} and {Tenti, M.} and {Viel, M.} and {Wiesmann, M.} and {Akrami, Y.} and {Allevato, V.} and {Anselmi, S.} and {Aubourg, E.} and {Ballardini, M.} and {Bertacca, D.} and {Bethermin, M.} and {Blanchard, A.} and {Blot, L.} and {Borgani, S.} and {Borlaff, A. S.} and {Bruton, S.} and {Cabanac, R.} and {Calabro, A.} and {Calderone, G.} and {Canas-Herrera, G.} and {Cappi, A.} and {Carvalho, C. S.} and {Castro, T.} and {Chambers, K. C.} and {Chary, R.} and {Contarini, S.} and {Cooray, A. R.} and {Cordes, O.} and {Costanzi, M.} and {Cucciati, O.} and {Davini, S.} and {De Caro, B.} and {Desprez, G.} and {D\'{\i}az-S\'anchez, A.} and {Di Domizio, S.} and {Dole, H.} and {Escoffier, S.} and {Ferrari, A. G.} and {Ferreira, P. G.} and {Ferrero, I.} and {Finoguenov, A.} and {Fontana, A.} and {Fornari, F.} and {Gabarra, L.} and {Ganga, K.} and {Garc\'{\i}a-Bellido, J.} and {Gautard, V.} and {Gaztanaga, E.} and {Giacomini, F.} and {Gianotti, F.} and {Gozaliasl, G.} and {Gregorio, A.} and {Hall, A.} and {Hartley, W. G.} and {Hildebrandt, H.} and {Hjorth, J.} and {Huertas-Company, M.} and {Ilbert, O.} and {Jimenez Mu\~noz, A.} and {Joudaki, S.} and {Kajava, J. J. E.} and {Kansal, V.} and {Karagiannis, D.} and {Kirkpatrick, C. C.} and {Lacasa, F.} and {Le Graet, J.} and {Legrand, L.} and {Libet, G.} and {Loureiro, A.} and {Macias-Perez, J.} and {Magliocchetti, M.} and {Mancini, C.} and {Mannucci, F.} and {Maoli, R.} and {Martins, C. J. A. P.} and {Matthew, S.} and {Maurin, L.} and {McPartland, C. J. R.} and {Metcalf, R. B.} and {Migliaccio, M.} and {Miluzio, M.} and {Monaco, P.} and {Moretti, C.} and {Morgante, G.} and {Nadathur, S.} and {Walton, Nicholas A.} and {Odier, J.} and {Oguri, M.} and {Patrizii, L.} and {Popa, V.} and {Potter, D.} and {Pourtsidou, A.} and {Reimberg, P.} and {Risso, I.} and {Rocci, P.-F.} and {Rollins, R. P.} and {Sahl\'en, M.} and {Scarlata, C.} and {Schaye, J.} and {Schneider, A.} and {Schultheis, M.} and {Sereno, M.} and {Shankar, F.} and {Sikkema, G.} and {Silvestri, A.} and {Simon, P.} and {Spurio Mancini, A.} and {Stadel, J.} and {Stanford, S. A.} and {Tanidis, K.} and {Tao, C.} and {Tessore, N.} and {Testera, G.} and {Tewes, M.} and {Teyssier, R.} and {Toft, S.} and {Tosi, S.} and {Troja, A.} and {Tucci, M.} and {Valieri, C.} and {Valiviita, J.} and {Vergani, D.} and {Vernizzi, F.} and {Verza, G.} and {Vielzeuf, P.} and {Weaver, J. R.} and {Zalesky, L.} and {Zinchenko, I. A.} and {Archidiacono, M.} and {Atrio-Barandela, F.} and {Bouvard, T.} and {Caro, F.} and {Dimauro, P.} and {Duc, P.-A.} and {Fang, Y.} and {Ferguson, A. M. N.} and {Gasparetto, T.} and {Gutierrez, C. M.} and {Kovaci\'{}c, I.} and {Kruk, S.} and {Le Brun, A. M. C.} and {Liaudat, T. I.} and {Montoro, A.} and {Mora, A.} and {Murray, C.} and {Pagano, L.} and {Paoletti, D.} and {Sarpa, E.} and {Viitanen, A.} and {Lesgourgues, J.} and {Mart\'{\i}n-Fleitas, J.} and {Scott, D.}},
	title = {Euclid - II. The VIS instrument},
	DOI= "10.1051/0004-6361/202450996",
	url= "https://doi.org/10.1051/0004-6361/202450996",
	journal = {A\&A},
	year = 2025,
	volume = 697,
	pages = "A2",
}

@misc{boyea2026,
      title={Galactic tides and the outer density profile of the Sculptor and Ursa Minor dwarf spheroidals}, 
      author={Daniel A. Boyea and Julio F. Navarro and Jaclyn Jensen and Raphaël Errani},
      year={2026},
      eprint={2604.24853},
      archivePrefix={arXiv},
      primaryClass={astro-ph.GA},
      url={https://arxiv.org/abs/2604.24853}, 
}

@ARTICLE{willmer2018,
       author = {{Willmer}, Christopher N.~A.},
        title = "{The Absolute Magnitude of the Sun in Several Filters}",
      journal = {\apjs},
         year = 2018,
        month = jun,
       volume = {236},
       number = {2},
          eid = {47},
        pages = {47},
          doi = {10.3847/1538-4365/aabfdf},
archivePrefix = {arXiv},
       eprint = {1804.07788},
 primaryClass = {astro-ph.SR},
       adsurl = {https://ui.adsabs.harvard.edu/abs/2018ApJS..236...47W}
}

@ARTICLE{revaz2009,
       author = {{Revaz}, Y. and {Jablonka}, P. and {Sawala}, T. and {Hill}, V. and {Letarte}, B. and {Irwin}, M. and {Battaglia}, G. and {Helmi}, A. and {Shetrone}, M.~D. and {Tolstoy}, E. and {Venn}, K.~A.},
        title = "{The dynamical and chemical evolution of dwarf spheroidal galaxies}",
      journal = {\aap},
         year = 2009,
        month = jul,
       volume = {501},
       number = {1},
        pages = {189-206},
          doi = {10.1051/0004-6361/200911734},
archivePrefix = {arXiv},
       eprint = {0904.4636},
 primaryClass = {astro-ph.GA},
       adsurl = {https://ui.adsabs.harvard.edu/abs/2009A&A...501..189R}
}

@article{Read_2006,
   title={The tidal stripping of satellites},
   volume={366},
   ISSN={1365-2966},
   url={http://dx.doi.org/10.1111/j.1365-2966.2005.09861.x},
   DOI={10.1111/j.1365-2966.2005.09861.x},
   number={2},
   journal={MNRAS},
   publisher={Oxford University Press (OUP)},
   author={Read, J. I. and Wilkinson, M. I. and Evans, N. W. and Gilmore, G. and Kleyna, J. T.},
   year={2006},
   month=Feb, pages={429–437} }

@misc{santossantos2025,
      title={The unabridged satellite luminosity function of Milky Way-like galaxies in $\Lambda$CDM: the contribution of "orphan" satellites}, 
      author={Isabel Santos-Santos and Carlos Frenk and Julio Navarro and Shaun Cole and John Helly},
      year={2025},
      eprint={2410.19475},
      archivePrefix={arXiv},
      primaryClass={astro-ph.GA},
      url={https://arxiv.org/abs/2410.19475}, 
}

@ARTICLE{2023MNRAS,
       author = {{Feldmann}, Robert and {Quataert}, Eliot and {Faucher-Gigu{\`e}re}, Claude-Andr{\'e} and {Hopkins}, Philip F. and {{\c{C}}atmabacak}, Onur and {Kere{\v{s}}}, Du{\v{s}}an and {Bassini}, Luigi and {Bernardini}, Mauro and {Bullock}, James S. and {Cenci}, Elia and {Gensior}, Jindra and {Liang}, Lichen and {Moreno}, Jorge and {Wetzel}, Andrew},
        title = "{FIREbox: simulating galaxies at high dynamic range in a cosmological volume}",
      journal = {\mnras},
         year = 2023,
        month = jul,
       volume = {522},
       number = {3},
        pages = {3831-3860},
          doi = {10.1093/mnras/stad1205},
archivePrefix = {arXiv},
       eprint = {2205.15325},
 primaryClass = {astro-ph.GA},
       adsurl = {https://ui.adsabs.harvard.edu/abs/2023MNRAS.522.3831F},
       }

@article{Vera_Casanova_2022,
   title={Linking the brightest stellar streams with the accretion history of Milky Way like galaxies},
   volume={514},
   ISSN={1365-2966},
   url={http://dx.doi.org/10.1093/mnras/stac1636},
   DOI={10.1093/mnras/stac1636},
   number={4},
   journal={MNRAS},
   publisher={Oxford University Press (OUP)},
   author={Vera-Casanova, Alex and Gómez, Facundo A and Monachesi, Antonela and Gargiulo, Ignacio and Pallero, Diego and Grand, Robert J J and Marinacci, Federico and Pakmor, Rüdiger and Simpson, Christine M and Frenk, Carlos S and Morales, Gustavo},
   year={2022},
   month=June, pages={4898–4911} }

@ARTICLE{LIGTH,
       author = {{Trujillo}, Ignacio and {D'Onofrio}, Mauro and {Zaritsky}, Dennis and {Madrigal-Aguado}, Alberto and {Chamba}, Nushkia and {Golini}, Giulia and {Akhlaghi}, Mohammad and {Sharbaf}, Zahra and {Infante-Sainz}, Ra{\'u}l and {Rom{\'a}n}, Javier and {Morales-Socorro}, Carlos and {Sand}, David J. and {Martin}, Garreth},
        title = "{Introducing the LBT Imaging of Galactic Halos and Tidal Structures (LIGHTS) survey. A preview of the low surface brightness Universe to be unveiled by LSST}",
      journal = {\aap},
         year = 2021,
        month = oct,
       volume = {654},
          eid = {A40},
        pages = {A40},
          doi = {10.1051/0004-6361/202141603},
archivePrefix = {arXiv},
       eprint = {2109.07478},
 primaryClass = {astro-ph.GA},
       adsurl = {https://ui.adsabs.harvard.edu/abs/2021A&A...654A..40T}
}

@misc{RomanLSB,
      title={Optimizing Roman's High Latitude Wide Area Survey for Low Surface Brightness Astronomy}, 
      author={Mireia Montes and Francesca Annibali and Michele Bellazzini and Alejandro S. Borlaff and Sarah Brough and Fernando Buitrago and Nushkia Chamba and Chris Collins and Ian Dell'Antonio and Ivanna Escala and Anthony H. Gonzalez and Benne Holwerda and Sugata Kaviraj and Johan Knapen and Anton Koekemoer and Seppo Laine and Pamela Marcum and Garreth Martin and David Martinez-Delgado and Chris Mihos and Massimo Ricotti and Ignacio Trujillo and Aaron E. Watkins},
      year={2023},
      eprint={2306.09414},
      archivePrefix={arXiv},
      primaryClass={astro-ph.GA},
      url={https://arxiv.org/abs/2306.09414}, 
}

@article{Simpson_2018,
   title={Quenching and ram pressure stripping of simulated Milky Way satellite galaxies},
   volume={478},
   ISSN={1365-2966},
   url={http://dx.doi.org/10.1093/mnras/sty774},
   DOI={10.1093/mnras/sty774},
   number={1},
   journal={MNRAS},
   publisher={Oxford University Press (OUP)},
   author={Simpson, Christine M and Grand, Robert J J and Gómez, Facundo A and Marinacci, Federico and Pakmor, Rüdiger and Springel, Volker and Campbell, David J R and Frenk, Carlos S},
   year={2018},
   month=Mar, pages={548–567} }

@article{Pakmor_2025,
    author = {Pakmor, Rüdiger and Fragkoudi, Francesca and Grand, Robert J J and Simpson, Christine M and Gómez, Facundo A and van de Voort, Freeke and Bieri, Rebekka and Trick, Wilma and Werhahn, Maria and Talbot, Rosie Y},
    title = {Auriga Superstars: Improving the resolution and fidelity of stellar dynamics in cosmological galaxy simulations},
    journal = {MNRAS},
    volume = {543},
    number = {4},
    pages = {4355-4368},
    year = {2025},
    month = {11},
    issn = {0035-8711},
    doi = {10.1093/mnras/staf1755},
    url = {https://doi.org/10.1093/mnras/staf1755},
    eprint = {https://academic.oup.com/mnras/article-pdf/543/4/4355/64633933/staf1755.pdf},
}

@ARTICLE{Cooper_2017,
       author = {{Cooper}, Andrew P. and {Cole}, Shaun and {Frenk}, Carlos S. and {Le Bret}, Theo and {Pontzen}, Andrew},
        title = "{Comparing semi-analytic particle tagging and hydrodynamical simulations of the Milky Way's stellar halo}",
      journal = {\mnras},
         year = 2017,
        month = aug,
       volume = {469},
       number = {2},
        pages = {1691-1712},
          doi = {10.1093/mnras/stx955},
archivePrefix = {arXiv},
       eprint = {1611.03497},
 primaryClass = {astro-ph.GA},
       adsurl = {https://ui.adsabs.harvard.edu/abs/2017MNRAS.469.1691C}
}

@ARTICLE{Johnston_2008,
       author = {{Johnston}, Kathryn V. and {Bullock}, James S. and {Sharma}, Sanjib and {Font}, Andreea and {Robertson}, Brant E. and {Leitner}, Samuel N.},
        title = "{Tracing Galaxy Formation with Stellar Halos. II. Relating Substructure in Phase and Abundance Space to Accretion Histories}",
      journal = {\apj},
         year = 2008,
        month = dec,
       volume = {689},
       number = {2},
        pages = {936-957},
          doi = {10.1086/592228},
archivePrefix = {arXiv},
       eprint = {0807.3911},
 primaryClass = {astro-ph},
       adsurl = {https://ui.adsabs.harvard.edu/abs/2008ApJ...689..936J}
}

@ARTICLE{Hendel_Johnston_2015,
       author = {{Hendel}, David and {Johnston}, Kathryn V.},
        title = "{Tidal debris morphology and the orbits of satellite galaxies}",
      journal = {\mnras},
         year = 2015,
        month = dec,
       volume = {454},
       number = {3},
        pages = {2472-2485},
          doi = {10.1093/mnras/stv2035},
archivePrefix = {arXiv},
       eprint = {1509.06369},
 primaryClass = {astro-ph.GA},
       adsurl = {https://ui.adsabs.harvard.edu/abs/2015MNRAS.454.2472H}
}

@article{de_Blok_2009,
   title={The Core‐Cusp Problem},
   volume={2010},
   ISSN={1687-7977},
   url={http://dx.doi.org/10.1155/2010/789293},
   DOI={10.1155/2010/789293},
   number={1},
   journal={Adv. Astron.},
   publisher={Wiley},
   author={de Blok, W. J. G.},
   editor={Brinks, Elias},
   year={2009},
   month=Nov }

@article{Pontzen_2012,
   title={How supernova feedback turns dark matter cusps into cores: Supernova feedback and dark matter cores},
   volume={421},
   ISSN={0035-8711},
   url={http://dx.doi.org/10.1111/j.1365-2966.2012.20571.x},
   DOI={10.1111/j.1365-2966.2012.20571.x},
   number={4},
   journal={MNRAS},
   publisher={Oxford University Press (OUP)},
   author={Pontzen, Andrew and Governato, Fabio},
   year={2012},
   month=Mar, pages={3464–3471} }

@ARTICLE{Penarubia_2010,
       author = {{Pe{\~n}arrubia}, Jorge and {Benson}, Andrew J. and {Walker}, Matthew G. and {Gilmore}, Gerard and {McConnachie}, Alan W. and {Mayer}, Lucio},
        title = "{The impact of dark matter cusps and cores on the satellite galaxy population around spiral galaxies}",
      journal = {\mnras},
         year = 2010,
        month = aug,
       volume = {406},
       number = {2},
        pages = {1290-1305},
          doi = {10.1111/j.1365-2966.2010.16762.x},
archivePrefix = {arXiv},
       eprint = {1002.3376},
 primaryClass = {astro-ph.GA},
       adsurl = {https://ui.adsabs.harvard.edu/abs/2010MNRAS.406.1290P}
}

@article{Akins_2021,
   title={Quenching Timescales of Dwarf Satellites around Milky Way–mass Hosts},
   volume={909},
   ISSN={1538-4357},
   url={http://dx.doi.org/10.3847/1538-4357/abe2ab},
   DOI={10.3847/1538-4357/abe2ab},
   number={2},
   journal={ApJ},
   publisher={American Astronomical Society},
   author={Akins, Hollis B. and Christensen, Charlotte R. and Brooks, Alyson M. and Munshi, Ferah and Applebaum, Elaad and Engelhardt, Anna and Chamberland, Lucas},
   year={2021},
   month=Mar, pages={139} }

@article{Rodr_guez_Cardoso_2025,
   title={The <i>AGORA</i> High-Resolution Galaxy Simulations Comparison Project: VII. Satellite quenching in zoom-in simulation of a Milky Way-mass halo},
   volume={698},
   ISSN={1432-0746},
   url={http://dx.doi.org/10.1051/0004-6361/202453639},
   DOI={10.1051/0004-6361/202453639},
   journal = {\aap},
   publisher={EDP Sciences},
   author={Rodríguez-Cardoso, Ramón and Roca-Fàbrega, Santi and Jung, Minyong and Nguyễn, Thịnh H. and Kim, Ji-hoon and Primack, Joel and Agertz, Oscar and Barrow, Kirk S. S. and Gallego, Jesus and Nagamine, Kentaro and Powell, Johnny W. and Revaz, Yves and Velázquez, Hector and Genina, Anna and Kim, Hyeonyong and Lupi, Alessandro and Abel, Tom and Cen, Renyue and Ceverino, Daniel and Dekel, Avishai and Oh, Boon Kiat and Quinn, Thomas R. and },
   year={2025},
   month=June, pages={A303} }

@ARTICLE{Miyamoto_Nagai_1975,
       author = {{Miyamoto}, M. and {Nagai}, R.},
        title = "{Three-dimensional models for the distribution of mass in galaxies}",
      journal = {PASJ},
         year = 1975,
       volume = {27},
        pages = {533--543},
       adsurl = {https://ui.adsabs.harvard.edu/abs/1975PASJ...27..533M}
}

@ARTICLE{2017MNRAS.468.4887W,
       author = {{Wang}, M.-Y. and {Fattahi}, Azadeh and {Cooper}, Andrew P. and {Sawala}, Till and {Strigari}, Louis E. and {Frenk}, Carlos S. and {Navarro}, Julio F. and {Oman}, Kyle and {Schaller}, Matthieu},
        title = "{Tidal features of classical Milky Way satellites in a {\ensuremath{\Lambda}} cold dark matter universe}",
      journal = {\mnras},
         year = 2017,
        month = jul,
       volume = {468},
       number = {4},
        pages = {4887-4901},
          doi = {10.1093/mnras/stx742},
archivePrefix = {arXiv},
       eprint = {1611.00778},
 primaryClass = {astro-ph.GA},
       adsurl = {https://ui.adsabs.harvard.edu/abs/2017MNRAS.468.4887W}
}

@ARTICLE{2025MNRAS.544...36C,
       author = {{Chiang}, Barry T. and {van den Bosch}, Frank C. and {Schive}, Hsi-Yu},
        title = "{The tidal evolution of anisotropic subhaloes: a new pathway to creating isotropic and cored satellites}",
      journal = {\mnras},
         year = 2025,
        month = nov,
       volume = {544},
       number = {1},
        pages = {36-52},
          doi = {10.1093/mnras/staf1639},
archivePrefix = {arXiv},
       eprint = {2411.03192},
 primaryClass = {astro-ph.GA},
       adsurl = {https://ui.adsabs.harvard.edu/abs/2025MNRAS.544...36C}
}

@ARTICLE{Delgado,
       author = {{Mart{\'\i}nez-Delgado}, David and {Cooper}, Andrew P. and {Rom{\'a}n}, Javier and {Pillepich}, Annalisa and {Erkal}, Denis and {Pearson}, Sarah and {Moustakas}, John and {Laporte}, Chervin F.~P. and {Laine}, Seppo and {Akhlaghi}, Mohammad and {Lang}, Dustin and {Makarov}, Dmitry and {Borlaff}, Alejandro S. and {Donatiello}, Giuseppe and {Pearson}, William J. and {Mir{\'o}-Carretero}, Juan and {Cuillandre}, Jean-Charles and {Dom{\'\i}nguez}, Helena and {Roca-F{\`a}brega}, Santi and {Frenk}, Carlos S. and {Schmidt}, Judy and {G{\'o}mez-Flechoso}, Mar{\'\i}a A. and {Guzman}, Rafael and {Libeskind}, Noam I. and {Dey}, Arjun and {Weaver}, Benjamin A. and {Schlegel}, David and {Myers}, Adam D. and {Valdes}, Frank G.},
        title = "{Hidden depths in the local Universe: The Stellar Stream Legacy Survey}",
      journal = {\aap},
         year = 2023,
        month = mar,
       volume = {671},
          eid = {A141},
        pages = {A141},
          doi = {10.1051/0004-6361/202245011},
archivePrefix = {arXiv},
       eprint = {2104.06071},
 primaryClass = {astro-ph.GA},
       adsurl = {https://ui.adsabs.harvard.edu/abs/2023A&A...671A.141M}
}

@ARTICLE{2016MNRAS.456.1359F,
       author = {{Fliri}, J{\"u}rgen and {Trujillo}, Ignacio},
        title = "{The IAC Stripe 82 Legacy Project: a wide-area survey for faint surface brightness astronomy}",
      journal = {\mnras},
         year = 2016,
        month = feb,
       volume = {456},
       number = {2},
        pages = {1359-1373},
          doi = {10.1093/mnras/stv2686},
archivePrefix = {arXiv},
       eprint = {1603.04474},
 primaryClass = {astro-ph.GA},
       adsurl = {https://ui.adsabs.harvard.edu/abs/2016MNRAS.456.1359F}
}

@ARTICLE{2010AJ....140..962M,
       author = {{Mart{\'\i}nez-Delgado}, David and {Gabany}, R. Jay and {Crawford}, Ken and {Zibetti}, Stefano and {Majewski}, Steven R. and {Rix}, Hans-Walter and {Fliri}, J{\"u}rgen and {Carballo-Bello}, Julio A. and {Bardalez-Gagliuffi}, Daniella C. and {Pe{\~n}arrubia}, Jorge and {Chonis}, Taylor S. and {Madore}, Barry and {Trujillo}, Ignacio and {Schirmer}, Mischa and {McDavid}, David A.},
        title = "{Stellar Tidal Streams in Spiral Galaxies of the Local Volume: A Pilot Survey with Modest Aperture Telescopes}",
      journal = {\aj},
         year = 2010,
        month = oct,
       volume = {140},
       number = {4},
        pages = {962-967},
          doi = {10.1088/0004-6256/140/4/962},
archivePrefix = {arXiv},
       eprint = {1003.4860},
 primaryClass = {astro-ph.CO},
       adsurl = {https://ui.adsabs.harvard.edu/abs/2010AJ....140..962M}
}

@ARTICLE{2001ApJ...555..504W,
       author = {{Wyithe}, J.~S.~B. and {Turner}, E.~L. and {Spergel}, D.~N.},
        title = "{Gravitational Lens Statistics for Generalized NFW Profiles: Parameter Degeneracy and Implications for Self-Interacting Cold Dark Matter}",
      journal = {\apj},
         year = 2001,
        month = jul,
       volume = {555},
       number = {1},
        pages = {504-523},
          doi = {10.1086/321437},
archivePrefix = {arXiv},
       eprint = {astro-ph/0007354},
 primaryClass = {astro-ph},
       adsurl = {https://ui.adsabs.harvard.edu/abs/2001ApJ...555..504W}
}
\let\cleardoublepage\clearpage

\begin{appendix}

\section{Additional progenitor models}\label{apx:additional_models}

In this appendix, we examine the sensitivity of our results to the initial size of the stellar component. We repeated a subset of the simulations using more extended Plummer profiles. The halo parameters, stellar masses and metallicities were kept unchanged with respect to the fiducial models to isolate the effect of increasing the stellar scale radius.

\begin{figure}
    \centering
    \includegraphics[width=0.85\linewidth]{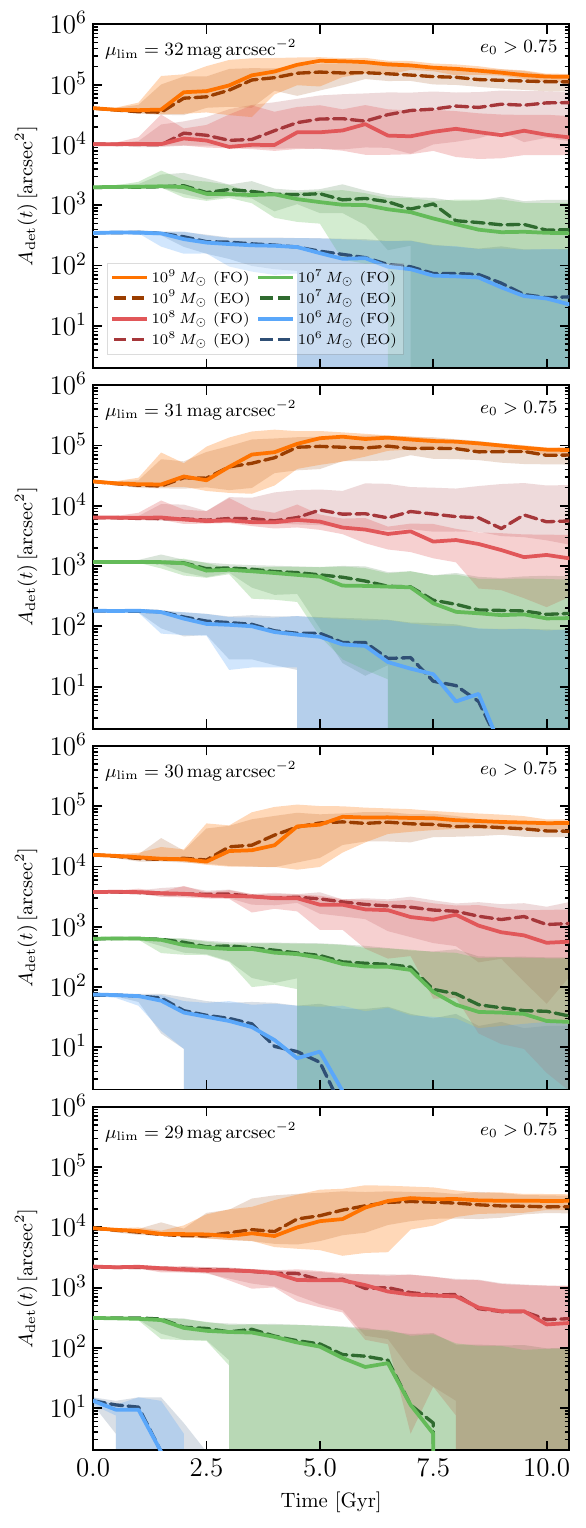}
    \caption{
    Time evolution of the detectable progenitor-light area for the extended cuspy progenitor models, restricted to orbits with initial eccentricity \(e_0 > 0.75\). Solid and dashed lines correspond to face-on and edge-on projections, respectively. For each stellar mass, the curves show the median evolution across the selected orbits, while the shaded regions indicate the 16th--84th percentile range at each time step. Panels correspond to different surface brightness detection limits, \(\mu_{\rm lim} = 32, 31, 30,\) and \(29\,\mathrm{mag\,arcsec^{-2}}\).
    }
    \label{fig:Times_extra}
\end{figure}

Figure~\ref{fig:Times_extra} shows that increasing the initial stellar extent enhances tidal disruption. For the lowest-mass progenitors, \(M_\star=10^6\) and \(10^7\,M_\odot\), the detectable area decreases strongly with time because the stellar component is stripped more efficiently and the resulting debris becomes too diffuse to remain above the adopted surface brightness limits. A similar behaviour appears for the \(M_\star=10^8\,M_\odot\) models at \(\mu_{\rm lim}=31\), 30, and \(29\,\mathrm{mag\,arcsec^{-2}}\). At the deepest limit, \(\mu_{\rm lim}=32\,\mathrm{mag\,arcsec^{-2}}\), the detectable area can increase at early times, when extended stream-like structures form, but it decreases again at later times as the debris becomes more phase-mixed and its surface brightness drops. The \(M_\star=10^9\,M_\odot\) models show a behaviour closer to the fiducial case, partly because their stellar scale radius is changed less substantially.

This experiment shows that the initial stellar size affects the normalisation and time evolution of \(A_{\rm det}\), especially for low- and intermediate-mass progenitors. Nevertheless, the qualitative trends found in the fiducial suite remain unchanged: detectable debris is favoured for massive progenitors, deep surface brightness limits, and eccentric orbits.

\section{Initial orbital parameters}
\label{apx:orbital_parameters}
Table~\ref{tab:objects} lists the initial positions, velocities, galactocentric distances, and eccentricity proxies of the 49 orbital configurations considered in this work.

\begin{table*}
\centering
\caption{Initial orbital parameters of the 49 simulated trajectories.}
\label{tab:objects}
\begin{tabular}{r r r r r r r r r}
\toprule
ID & \(e_0\) & \(r_0\) [kpc] & \(x_0\) [kpc] & \(y_0\) [kpc] & \(z_0\) [kpc] &
\(v_{x,0}\) [km s\(^{-1}\)] & \(v_{y,0}\) [km s\(^{-1}\)] & \(v_{z,0}\) [km s\(^{-1}\)] \\
\midrule
0  & 0.9896 & 174.7 & 3.8    & -85.1  & 152.6  & 15.3   & 0.7    & 9.5 \\
1  & 0.9804 & 245.9 & -194.3 & 34.0   & -146.8 & -2.6   & -14.9  & -13.4 \\
2  & 0.9639 & 185.5 & -99.8  & 34.8   & 152.5  & -9.1   & -26.2  & 14.7 \\
3  & 0.9631 & 110.3 & -58.0  & -82.2  & -45.1  & 19.8   & -13.9  & 32.1 \\
4  & 0.9557 & 83.4  & -27.0  & 50.2   & 60.9   & -41.8  & -22.5  & -8.2 \\
5  & 0.9209 & 166.7 & 110.2  & 47.3   & 115.8  & -17.8  & 41.5   & 1.4 \\
6  & 0.9194 & 191.3 & -122.3 & 65.0   & 131.9  & -19.5  & -36.7  & -13.3 \\
7  & 0.9114 & 227.2 & 223.6  & 12.2   & 38.5   & -2.2   & 40.0   & 8.8 \\
8  & 0.9080 & 206.4 & -47.9  & -139.4 & 144.5  & 41.4   & -14.2  & 1.5 \\
9  & 0.9058 & 201.3 & 161.6  & 53.8   & 107.3  & -14.1  & 42.4   & -5.3 \\
10 & 0.9028 & 100.9 & 87.6   & 49.7   & 6.0    & -31.1  & 54.9   & -12.6 \\
11 & 0.8995 & 110.3 & 15.9   & 29.7   & -105.0 & -55.2  & 29.6   & 6.6 \\
12 & 0.8950 & 108.9 & 93.5   & -48.8  & 27.0   & 29.8   & 57.1   & -2.2 \\
13 & 0.8926 & 130.2 & -1.5   & -128.0 & 23.7   & 59.5   & -0.7   & -3.4 \\
14 & 0.8904 & 184.7 & 127.1  & 31.5   & 130.3  & -12.1  & 48.7   & -8.9 \\
15 & 0.8874 & 150.3 & -3.1   & 3.5    & 150.2  & -42.4  & -37.8  & -9.4 \\
16 & 0.8837 & 227.3 & -6.2   & -48.3  & 222.1  & 46.7   & -6.0   & 25.5 \\
17 & 0.8823 & 133.5 & -75.0  & 100.1  & -46.5  & -48.1  & -36.1  & -14.2 \\
18 & 0.8817 & 224.1 & -10.3  & -223.6 & 10.1   & 45.8   & -2.1   & -13.1 \\
19 & 0.8646 & 112.0 & -95.2  & -49.8  & -31.5  & 33.2   & -63.5  & -8.2 \\
20 & 0.8503 & 99.1  & -41.5  & 60.7   & -66.5  & -64.9  & -44.4  & 24.6 \\
21 & 0.8438 & 102.6 & 13.5   & 21.2   & -99.5  & -68.2  & 43.5   & 9.6 \\
22 & 0.8225 & 160.2 & 135.2  & -78.7  & 34.6   & 34.6   & 59.4   & 6.6 \\
23 & 0.7979 & 170.7 & 122.8  & 97.4   & 67.8   & -44.3  & 55.9   & 0.7 \\
24 & 0.7715 & 111.7 & 4.9    & -41.6  & 103.6  & 93.1   & 11.0   & -22.0 \\
25 & 0.7670 & 159.4 & 90.2   & 12.6   & -130.9 & -10.9  & 78.0   & 18.9 \\
26 & 0.7446 & 125.8 & -59.9  & -102.9 & 40.5   & 75.1   & -43.7  & 36.7 \\
27 & 0.7295 & 185.5 & 63.7   & 50.0   & 166.9  & -48.8  & 62.1   & -31.3 \\
28 & 0.7182 & 112.8 & 20.3   & -39.7  & -103.6 & 92.3   & 47.2   & -1.9 \\
29 & 0.6967 & 129.7 & -8.5   & -113.8 & -61.6  & 99.9   & -7.5   & -4.7 \\
30 & 0.6953 & 115.4 & 42.8   & -16.1  & 106.0  & 37.4   & 99.8   & -16.0 \\
31 & 0.6750 & 241.1 & 133.6  & 26.4   & -199.0 & -14.5  & 73.5   & 35.6 \\
32 & 0.6714 & 204.8 & -47.0  & -89.7  & 178.0  & 73.6   & -38.6  & -2.4 \\
33 & 0.6703 & 138.2 & -101.2 & 49.2   & -80.2  & -43.6  & -89.8  & -22.5 \\
34 & 0.6674 & 216.5 & 105.4  & 24.0   & 187.6  & -18.1  & 79.2   & 7.7 \\
35 & 0.6572 & 201.2 & -24.4  & -195.9 & 38.8   & 84.9   & -10.6  & 2.3 \\
36 & 0.6561 & 171.7 & 69.8   & -104.6 & 116.9  & 76.9   & 51.3   & -13.8 \\
37 & 0.6534 & 127.6 & -29.7  & -9.5   & -123.7 & 33.0   & -103.0 & 10.8 \\
38 & 0.6411 & 196.3 & 43.7   & -40.1  & -187.1 & 60.6   & 66.0   & 29.9 \\
39 & 0.6131 & 87.3  & 69.5   & -16.3  & -50.2  & 31.4   & 134.3  & 8.7 \\
40 & 0.6129 & 143.9 & -20.8  & 4.8    & 142.3  & -24.3  & -105.1 & 14.6 \\
41 & 0.6107 & 232.8 & 98.4   & -50.1  & 204.9  & 38.5   & 75.6   & -8.8 \\
42 & 0.5712 & 100.7 & -19.9  & 82.2   & -54.6  & -130.6 & -31.6  & 20.5 \\
43 & 0.5377 & 249.2 & -121.9 & -203.1 & -77.3  & 75.6   & -45.4  & 15.6 \\
44 & 0.5332 & 247.0 & 181.7  & 29.2   & -164.7 & -14.3  & 88.9   & 8.1 \\
45 & 0.5162 & 176.1 & -154.9 & 73.2   & 40.6   & -46.2  & -97.6  & 12.9 \\
46 & 0.5094 & 158.4 & 86.1   & -89.7  & 98.1   & 83.3   & 79.9   & 3.6 \\
47 & 0.4574 & 168.2 & -126.2 & -56.4  & -95.8  & 47.7   & -106.8 & 19.9 \\
48 & 0.4453 & 144.8 & 84.4   & -62.2  & 99.8   & 76.2   & 103.3  & 2.0 \\
\bottomrule
\end{tabular}
\tablefoot{
Positions and distances are given in kpc, and velocities in km\,s$^{-1}$. The eccentricity proxy $e_0$ is defined in Sect.~\ref{sec:orbits}.
}

\end{table*}

\end{appendix}

\end{document}